%% file: 0-facct.tex
\documentclass[manuscript,screen,authorversion,nonacm]{acmart}

\usepackage{tcolorbox}
\usepackage{xcolor}
 \definecolor{antiquefuchsia}{rgb}{0.57, 0.36, 0.51}

\AtBeginDocument{%
  \providecommand\BibTeX{{%
    \normalfont B\kern-0.5em{\scshape i\kern-0.25em b}\kern-0.8em\TeX}}}

\setcopyright{acmlicensed}
\copyrightyear{2024}
\acmYear{2024}
\acmDOI{XXXXXXX.XXXXXXX}

\acmConference[FAccT '24]{Make sure to enter the correct
  conference title from your rights confirmation emai}{June 03--06,
  2024}{Rio de Janeiro, Brazil}
\acmISBN{978-1-4503-XXXX-X/18/06}

\acmSubmissionID{421}

\begin{document}

\title[A Multiple-Stakeholder Ethics Discussion on LLMs in Computing Education]{``The teachers are confused as well'': A Multiple-Stakeholder Ethics Discussion on Large Language Models in Computing Education}
    
\author{Kyrie Zhixuan Zhou}
\thanks{Authors' addresses: Kyrie Zhixuan Zhou, zz78@illinois.edu, University of Illinois at Urbana-Champaign, Champaign, IL, United States; Zachary Kilhoffer, dzk2@illinois.edu, University of Illinois at Urbana-Champaign, Champaign, IL, United States; Madelyn Rose Sanfilippo, madelyns@illinois.edu, University of Illinois at Urbana-Champaign, Champaign, IL, United States; Ted Underwood, tunder@illinois.edu, University of Illinois at Urbana-Champaign, Champaign, IL, United States; Ece Gumusel, egumusel@iu.edu, Indiana University Bloomington, Bloomington, IN, United States; Mengyi Wei, mengyi.wei@tum.de, Technical University of Munich, Munich, Germany; Abhinav Choudhry, ac62@illinois.edu, University of Illinois at Urbana-Champaign, Champaign, IL, United States; Jinjun Xiong, jinjun@buffalo.edu, University at Buffalo, Buffalo, NY, United States.}
\email{zz78@illinois.edu}
\affiliation{
  \institution{University of Illinois at Urbana-Champaign}
  \city{Champaign}
  \state{Illinois}
  \country{United States}
}

\author{Zachary Kilhoffer}
\email{dzk2@illinois.edu}
\affiliation{
  \institution{University of Illinois at Urbana-Champaign}
  \city{Champaign}
  \state{Illinois}
  \country{United States}
}

\author{Madelyn Rose Sanfilippo}
\email{madelyns@illinois.edu}
\affiliation{
  \institution{University of Illinois at Urbana-Champaign}
  \city{Champaign}
  \state{Illinois}
  \country{United States}
}

\author{Ted Underwood}
\email{tunder@illinois.edu}
\affiliation{
  \institution{University of Illinois at Urbana-Champaign}
  \city{Champaign}
  \state{Illinois}
  \country{United States}
}

\author{Ece Gumusel}
\email{egumusel@iu.edu}
\affiliation{
  \institution{Indiana University Bloomington}
  \city{Bloomington}
  \state{Indiana}
  \country{United States}
}

\author{Mengyi Wei}
\email{mengyi.wei@tum.de}
\affiliation{
  \institution{Technical University of Munich}
  \city{Munich}
  \state{}
  \country{Germany}
}

\author{Abhinav Choudhry}
\email{ac62@illinois.edu}
\affiliation{
  \institution{University of Illinois at Urbana-Champaign}
  \city{Champaign}
  \state{Illinois}
  \country{United States}
}

\author{Jinjun Xiong}
\email{jinjun@buffalo.edu}
\affiliation{
  \institution{University at Buffalo}
  \city{Buffalo}
  \state{New York}
  \country{United States}
}

\renewcommand{\shortauthors}{Zhou and Kilhoffer, et al.}

\begin{abstract}
\noindent Large Language Models (LLMs) are advancing quickly and impacting people's lives for better or worse. In higher education, concerns have emerged such as students' misuse of LLMs and degraded education outcomes. To unpack the ethical concerns of LLMs for higher education, we conducted a case study consisting of stakeholder interviews (n=20) in higher education computer science. We found that students use several distinct mental models to interact with LLMs - LLMs serve as a tool for (a) writing, (b) coding, and (c) information retrieval, which differ somewhat in ethical considerations. Students and teachers brought up ethical issues that directly impact them, such as inaccurate LLM responses, hallucinations, biases, privacy leakage, and academic integrity issues.
% and degraded human autonomy. 
Participants emphasized the necessity of guidance and rules for the use of LLMs in higher education, including teaching digital literacy, rethinking education, and having cautious and contextual policies. We reflect on the ethical challenges and propose solutions. 
\end{abstract}


% \ccsdesc[500]{Security and privacy~Social aspects of security and privacy}


% \received{20 February 2007}
% \received[revised]{12 March 2009}
% \received[accepted]{5 June 2009}

\maketitle

\input{1-introduction}
\input{2-related}
\input{3-method}
\input{4-results}
\input{5-discussion}
\input{6-conclusion}

% \begin{acks}
% To Robert, for the bagels and explaining CMYK and color spaces.
% \end{acks}

\newpage
\bibliographystyle{ACM-Reference-Format}
\bibliography{7-references}

\appendix

\input{8-interview}
\input{9-demographic}

\end{document}

%% file: 1-introduction.tex
\section{Introduction}

Large Language Models (LLMs) have advanced significantly in recent years, leading to widespread adoption of the powerful products they enable. As of 2024, millions of people use LLM-powered chatbots like OpenAI's ChatGPT and Baidu's Ernie for a variety of purposes in diverse contexts \cite{zhao2023survey}. LLMs can interpret and generate natural and computer languages to assist cognitively demanding tasks, offering creativity support \cite{di2022idea}, task automation \cite{wen2023empowering}, and much more. 

Education is rich with LLM use cases, leading to an emerging body of literature on LLMs in higher education \cite{laato2023ai, ansari2023mapping, aithal2023application, perera2023ai, wang2023navigating}. Research suggests LLMs help engage students, facilitate collaboration, and personalize learning experiences \cite{cotton2023chatting}. Researchers have highlighted these and other benefits, as well as risks, of LLMs in education \cite{hasanein2023drivers, eager2023prompting, dempere2023impact, mvondo2023generative, gill2024transformative}.

The ethics of LLMs in Computer Science (CS) education (and related technical/coding disciplines) are notable for several reasons. First, many CS students will work on LLMs and related products and serve as decision-makers in various industries. Second, CS students can make use of two very important features LLMs offer: writing text and coding. This is a profound innovation with significant effects on coding education. Traditionally, early CS classes entail a focus on coding assignments, but LLMs offer a way to generate custom code without learning to do so, comprehending its meaning, or relying on other people. Third, CS students are better positioned to understand how LLMs function, which may have interesting impacts on their LLM usage and ethical concerns.

% A lively debate is ongoing about the ethics of LLMs in education, with concerns including privacy \cite{zhang2023s}, biased LLM behavior \cite{felkner2023winoqueer}, academic integrity \cite{singh2023exploring}, and user awareness of LLM limitations \cite{lau2023ban}. 

Existing research on LLMs in CS education concerns both practical and ethical concerns. 
% Researchers have proposed further work on several points concerning the ethics of LLMs in CS education. 
For example, it remains unclear if and how novice programmers validate LLM outputs \cite{lau2023ban}, which is essential for effective usage. Lau and Guo therefore emphasized the need to build a theory of how AI novices construct mental models on LLMs. Existing research is contradictory on how LLMs impact CS students' critical thinking and problem-solving \cite{halaweh2023chatgpt, kasneci2023chatgpt, singh2023exploring}. Further gaps include: how CS curricula can use LLMs toward greater equity and access, rather than furthering a digital divide \cite{lau2023ban}; how introductory CS curricula and assessment thereof may evolve, given the ease of doing certain coursework with LLMs \cite{lau2023ban}; and what type of policies are desirable to govern LLM usage in CS education \cite{singh2023exploring}.

To bridge these research gaps, we engaged 20 CS education stakeholders in the ethical discussion with in-depth interviews. Participants shared their experiences and opinions on LLMs in CS education. Stakeholders held nuanced and intertwined ethical concerns and overwhelmingly supported explicit but permissive LLM policies.  
% A tiered regulation system engaging students, teachers, and university departments was proposed as a way to regulate LLM use in higher education.
% For example, students' privacy concerns interact with academic integrity  (``teachers will know I use ChatGPT in assignments if OpenAI discloses my ChatGPT use''). Instructors suggested course-level guidance on LLM use was effective in nudging the students away from ``over-using'' LLMs in coursework. Other examples include context-sensitive education policy, incorporating ethical LLM usage within mandatory digital literacy education, and adjustments to educational outcome assessment.
Our main contributions are three-fold. First, we provide timely empirical evidence on LLMs in CS education, elaborating and refining themes, including privacy and coders' responsibility to understand LLM limitations. 
% \zakcomment{Additionally, our evidence surfaced a novel concept - coders' responsibility to anticipate and proactively mitigate LLM harms. We also refine the theme of privacy, elaborating how privacy risks differently impact multiple parties.}
Second, we propose that people use three main mental models for LLMs in CS education -- coding tool, writing tool, and information tool --
and reflect on the ethical implications. 
Third, we add to the policy and governance discussion by presenting and assessing stakeholders' suggestions for policies to support ethical and effective LLM usage in educational settings.

%% file: 2-related.tex
%%%%%%%%%%%%%%%%%%%%%%%%%%%%%%%%%%%%%%%%%%%%%%%%%%%%%%
\section{Related Work}
We situate this work in two fields: (1) LLMs in higher education and (2) LLMs in CS education, including ethics and governance discussions.

\subsection{LLMs in Higher Education}
 
LLMs in education is an emerging HCI topic area \cite{zhou2023toward}. The literature is relatively immature; LLMs only made the leap to widely available and powerful chatbot products around November 2022, when ChatGPT was released \cite{fui2023generative}.
Notably, nearly all of the research that surfaced with the keywords \{LLM, large language model, generative AI\} concerned ChatGPT, suggesting its wide adoption and impact in the previous year. This area of research is closely related to broader fields of AI scholarship, spanning privacy/security education; FATE (fairness, accountability, transparency, and ethics in AI); and so on \cite{malinka2023educational, mischak2023challenges,orenstrakh2023detecting,singh2023exploring,felkner2023winoqueer,huang2023bias,alkaissi2023artificial,lau2023ban,gupta2023chatgpt,zhang2023s,george2023chatgpt,zhou2022moral,zhou2023m,kilhoffer2023ai}.
% Research on ethics mostly concerns academic integrity \cite{malinka2023educational, mischak2023challenges, orenstrakh2023detecting,singh2023exploring}, but also how to handle LLM bias in general \cite{felkner2023winoqueer}, biased code \cite{huang2023bias}, inaccuracy and hallucination \cite{alkaissi2023artificial,lau2023ban}, privacy implications \cite{gupta2023chatgpt,zhang2023s}, and human autonomy \cite{george2023chatgpt}.

Situated among broader studies of LLMs in education \cite{abdelghani_gpt-3-driven_2023, moore2023empowering, javaid2023unlocking}, researchers have considered LLMs in the higher education context \cite{laato2023ai, ansari2023mapping, aithal2023application, perera2023ai, wang2023navigating, vargas2023challenges, fuchs2023exploring, tajik2023comprehensive, dempere2023impact}. 
Researchers have used surveys and interviews to understand how students \cite{ngo2023perception, shaengchart2023factors, strzelecki2023use} and instructors \cite{kiryakova2023chatgpt, baskara2023chatgpt} use LLMs in various use cases. Other topics include: benchmarking LLM capabilities vs. each other and humans \cite{rudolph2023war, newton2023chatgpt}; optimization strategies (e.g., prompt engineering \cite{lo2023clear}); ethical challenges \cite{huallpa2023exploring}; and LLM governance/policy \cite{fowler2023first}.

\textbf{Learning with LLMs.} Given the many use cases of LLMs in educational settings, both discovered and not, many studies investigated or proposed novel ways students use LLMs and the potential benefits. A systematic literature review revealed notable benefits of utilizing ChatGPT in higher education, including research support, and automated grading \cite{dempere2023impact}. For example, LLMs can serve as a \textbf{language tool}. LLMs were found to help international students cope with language anxieties by serving as ``conversational companions'' \cite{bao_can_nodate}. LLMs were found to promote English as a Foreign Language (EFL) learners’ willingness to communicate \cite{Tai_Chen_2023}. LLMs also seem to enhance curiosity-driven learning and serve to promote curiosity expression \cite{bao_can_nodate, Tai_Chen_2023}.
Students noted benefits of LLMs, such as time efficiency, convenient information access, personalized tutoring and feedback, and writing assistance \cite{ngo2023perception}. On the other hand, concerns about LLMs in higher education included plagiarism, digital literacy gaps, and AI-induced anxiety \cite{dempere2023impact}.
Students expressed concerns about LLM inaccuracy, such as hallucinated citations and incorrect idiom usage \cite{ngo2023perception}.

\textbf{Teaching with LLMs.} LLMs can assist teachers with many routine and creative tasks \cite{firaina2023exploring}. University instructors perceived LLMs as a means to support time-consuming teaching activities, stimulate critical thinking and creativity, and engage learners \cite{kiryakova2023chatgpt}. Early-career lecturers found LLMs beneficial to enhance their ability to construct complicated academic arguments, improve efficiency in research and teaching, and enhance critical thinking capabilities \cite{baskara2023chatgpt, firaina2023exploring}. LLMs were found useful to help formulate learning goals if provided high-quality prompts \cite{jacobsen2023promises}. Researchers have explored fine-tuning language models for crafting quality data science learning materials \cite{bhat_towards_2022}. Onal and Kulavuz-Onal found ChatGPT effective in writing assessment questions and tasks, provided that instructors carefully evaluate the reliability and accuracy of the LLM's output text \cite{onal2023cross}. Simultaneously, challenges emerged, such as initial technical hurdles, occasional incorrect outputs, and concerns about over-dependency \cite{baskara2023chatgpt, firaina2023exploring}. Professors were concerned that (1) unethical LLM use threatened the validity and fairness of assessment practices, and (2) learners would blindly trust LLMs without validating their text outputs, harming the acquisition of knowledge and skills.

% LLMs were better off as support for autonomous learning and less helpful in emotional support \cite{pinochet2023collaborative}. 

Overall, LLMs are a double-edged sword.
Education stakeholders including students, faculty, and education experts/leaders agreed that LLMs can be readily used or misused \cite{hasanein2023drivers}. Ethical harms, like plagiarism, and practical benefits, like supportive learning environments, are both foreseeable outcomes of further LLM use \cite{eager2023prompting}. 

\subsection{LLMs in Computer Science Education}

CS education has particularities that make it valuable to explore. For example, CS education often covers AI ethics as a topic \cite{kilhoffer2023technical}; classes often emphasize both coding and writing as part of learning and assessment \cite{dansdill2008exposing}. 
Lastly, there is a high likelihood that CS students will become tomorrow's AI architects and developers.

% Theoretical frameworks for enhancing the overall educational experience in teaching-learning activities within universities have been proposed (e.g., \cite{ilieva2023effects}). However, 
% Differentiating LLMs use between different disciplines is important since LLMs performed differently and had varying stakes \cite{baskara2023exploring}. For example, ChatGPT powered by GPT-3.5 performed below the average student in all written tasks in medical imaging \cite{currie2023chatgpt} -- errors in medicine are often lethal. In this paper, we contribute an ethics and policy discussion to the literature on LLMs in CS education \cite{albonico2023report, zheng2023chatgpt, banerjee2023understanding, lankes2023ai}.

\textbf{LLMs are valuable but imperfect tools for coding tasks.} LLMs can generate usable code upon receiving natural language prompts regarding code requirements \cite{liu2023your}, or produce natural language explanations from code prompts \cite{macneil_automatically_2022, macneil_automatically_2022} -- a valuable task if the intention is to learn about coding. 
Additional uses include resolving bugs \cite{biswas2023role}, improving code quality (e.g., fixing formatting issues, refactoring) \cite{backstrom2023code}, and even helping with complex software architecting tasks \cite{ahmad2023towards}. CS instructors could benefit from many applications, like using LLMs for grading programming tasks \cite{jukiewiczfuture}.

However, research suggests limitations for LLM-produced code, requiring verification from human coders \cite{liu2023your}. At least two studies found LLMs performed well in code generation but not software testing \cite{loubier2023chatgpt, jalil2023chatgpt}. Achieving good results for long/complex programming tasks required careful and repeated prompting. Otherwise, code inconsistencies and truncations caused problems \cite{qureshi2023exploring}. In light of these limitations, researchers have considered how to improve LLM results for coding tasks. Strategies in the literature have included iteratively prompting and validating responses \cite{rajala2023call}, prompt engineering \cite{gero_sparks_2022, qadir_engineering_2022}, and meta-prompting \cite{Reynolds_McDonell_2021}. Kumar et al. tested different instructor guidance strategies for using LLMs on coding tasks. The authors emphasized the nuance of interactions between undergrad CS students, instructors, and LLMs, suggesting that helpful and unhelpful interactions are both possible \cite{kumar2023impact}. Thus, instructors must provide considerable guidance for beginner coders to succeed in using LLMs as coding tools.

\textbf{Coding without groking.}
Researchers have considered how LLMs serve as a functional and/or educational coding tool.
Numerous studies suggest a tradeoff; LLMs may help students code better and more efficiently \cite{biswas2023role, backstrom2023code, ahmad2023towards}, but also inhibit their understanding of programming or AI concepts \cite{reiche2023bridging, arias2023comparing, wang2023exploring}. The issue seems related to the fact that LLMs allow ``doing coding'' without the cognitive effort required for understanding. One study suggests that current LLM debates echo earlier research on automated hints for CS education -- ultimately, educators must consider the balance between allowing LLMs for immediate coding help and considerations for long-term learning \cite{kumar2023impact}.

\textbf{LLMs for noobs.} Studies indicate that coding experience significantly impacts trust in and reliance on LLMs for coding tasks; in particular, LLMs facilitate coding but may harm beginners' efforts to learn coding.
Wang et al. interviewed CS instructors, who believed that 1st-year students often trust and use LLM-generated code with little consideration.
Instructors therefore felt concerned that ``students with underdeveloped mental models may experience shallow learning when using AI assistants'' \cite{wang2023exploring}. One industry study considered how software developers used LLMs for coding, finding that junior developers were more likely to trust and rely on LLM-generated code \cite{hornemalm2023chatgpt}. 
Laato et al. argue that over-reliance on ChatGPT for programming can hinder students’ learning of basic routine programming skills \cite{laato2023ai}. Further, Arias et al. separated first-year CS students into two groups -- one using ChatGPT and one using Google -- in an experiment comparing their performance on programming tasks \cite{arias2023comparing}. 
To that end, the ChatGPT group performed programming tasks better and faster. 
However, the group using Google performed better in a test of their comprehension. The authors observed that the ChatGPT group tended to copy and paste code rather than think, which hindered their learning. This evidence suggests LLMs are effective tools if the goal is to succeed in coding tasks, but overcoming challenges too easily may interfere with students' learning opportunities. This distinction is important if the goal is for students to understand and complete coding tasks on their own, as is more likely the case for early CS students. 
% A case study found that using LLMs helped business students complete machine learning projects, but did not deepen their understanding of coding \cite{reiche2023bridging}. 
% For business students, but presumably not CS students, this outcome may be acceptable.
% This is quite similar to an earlier issue, where math classes had to adapt to the pocket calculator \zakcomment{(cite)}. 
Still, overly restrictive guardrails are inadvisable as beginner coders also value LLMs for Q\&A, improving thinking skills, and building their confidence -- not merely generating code that ``works'' \cite{yilmaz2023augmented}.

% A formative study in an undergraduate CS classroom and a controlled experiment revealed that direct LLM answers marginally improved performance while refining student solutions fostered trust \cite{kumar2023impact}. 

% \zakcomment{Put additional evidence here, especially illustrating learning pros and cons of LLMs in CS edu.}

% Students were well aware of the benefits and risks of using LLMs in programming learning -- they expressed benefits such as providing fast and mostly correct answers to questions, improving thinking skills, facilitating debugging, and increasing self-confidence, as well as limitations such as getting students used to laziness, being unable to answer some questions or giving incomplete/incorrect answers, and causing professional anxiety in students \cite{yilmaz2023augmented}. 

% Hoq found that Abstract Syntax Tree-based deep learning models were effective in differentiating code written by a student or ChatGPT in an educational setting with accuracies higher than 90\% \cite{hoq2022detecting} -- the less than 10\% of inaccuracy, however, makes the detection models less trustworthy.

%%%%%%%%%%%%%%%%%%%%%%%%%%%%%%%%%%%%%%%%%%%%%%%%%%
\subsection{Ethics and Governance of LLMs in Computer Science Education} 

Many ethical topics arise with LLMs in CS education. Here we outline several important issues already identified.

\textbf{Academic integrity} issues have a long history in higher education and academia \cite{macfarlane2014academic} -- LLMs likely complicate the landscape. 
% Research considering the ethics of LLMs in CS education finds \zakcomment{(... tbd)}. Possibly ~\cite{choksi2023assessment,sanfilippo2023privacy}
Evidence thus far suggests LLMs have a complex, disruptive impact on assessment and other elements of academic integrity. 
This will likely require adapting curricula and/or governance strategies.
To start, LLMs can be easily used to cheat in CS courses \cite{malinka2023educational}. 
Traditional plagiarism checkers will not typically work because LLMs generate unique texts. 
Moreover, detecting LLM-generated content remains difficult, and existing solutions are inaccurate \cite{mischak2023challenges}. 
One study evaluated state-of-the-art LLM-content detectors, finding they generally perform poorly but are especially inaccurate with programming code and non-English text \cite{orenstrakh2023detecting}. 
Azoulay et al. proposed combining educational initiatives, plagiarism detection mechanisms, and watermarking techniques to regulate AI-produced code \cite{azoulay2023let}.

\textbf{Hallucination/inaccuracy.}
Currie et al. found that ChatGPT was prone to hallucinating information, posing a risk to professionalism, ethics, and integrity \cite{currie2023academic}. 
Multiple researchers expressed concern about equity within higher education; for example, if the free and paid versions of ChatGPT perform differently, students with fewer resources may fall behind in learning and assessment outcomes \cite{lau2023ban, bordt2023chatgpt}. 

\textbf{Bias} is prevalent in LLMs \cite{zhou2023public}, originating in training data and appearing in generated text \cite{liang_towards_2021,bender_dangers_2021}. 
LLM code also contains bias; code may contain offensive stereotypes in variable names or strings \cite{becker_programming_2023, bird_taking_2022, chen_evaluating_2021}, or privilege certain people over others (e.g., ``write code that decides if someone gets social security benefits'' \cite{huang_bias_2024}). 
Countermeasures may include carefully prompting ChatGPT or explicitly teaching students about bias.
 
\textbf{Privacy} is an important consideration as LLM users constantly encounter trade-offs between privacy, utility, and convenience \cite{zhang2023s}. In a student survey (n=400) in Thailand, Shaengchart et al. found that privacy and security were not significantly related to students' acceptance of ChatGPT in educational contexts \cite{shaengchart2023factors}. This finding could suggest that students lack awareness of LLM vulnerabilities. However, it is unclear how the survey asked about privacy and security and how the responses were operationalized. Recent research on applications of emerging technology in higher education, including video conferencing and AI-based academic integrity tools, are inadequately governed with respect to privacy under the auspices of the Family Educational Rights and Privacy Act (FERPA) ~\cite{cohney2021virtual,sanfilippo2023privacy}.
% Not only does FERPA fail to address the diverse data types associated with these technologies, but their control by commercial entities, rather than educational professionals challenges norms and expectations of privacy within educational contexts ~\cite{sanfilippo2023privacy}. 
Many states have sought to fill in the regulatory gaps left by FERPA, with 128 different laws in total~\cite{cohney2021virtual}. However, this leaves significant uncertainty and inconsistency in privacy protections concerning LLMs in educational settings.

%% file: 3-method.tex
%%%%%%%%%%%%%%%%%%%%%%%%%%%%%%%%%%%%%%%%%%%%%
\section{Methodology}
We conducted an interview study to explore the experiences of CS education stakeholders including students, professors, and AI professionals who have worked closely with LLMs. 
% We facilitated discussion with explicit ethical concerns and a examples of potential LLM policies toward informing more contextual education policies and preparing students for the future human-LLM collaborative workplace. 
We used individual interviews to protect students' privacy, as their LLM usage could violate course or university policies. 
% During the interviews, several students asked if the study was private, which was confirmed by the researchers. 

\subsection{Study Participants}
Our goal was understanding LLM use in higher education, specifically computing education. Thus, professors and students in computer science, information science, and other computing-related disciplines were target groups. Industrial practitioners' insights were considered to (1) hear from people with practical LLM/AI experience; and (2) prevent us from synthesizing education policies that were irrelevant or counter-productive for students' future careers. 

\textbf{Recruitment.} 
We recruited 4 undergraduate students (U1-U4), 4 master's students (M1-M4), 4 doctoral students (D1-D4), 4 professors (P1-P4), and 4 industrial practitioners (I1-I4) for interviews. The recruitment message was distributed on Twitter and to personal contacts of the researchers, followed by a snowball sampling approach. It conveyed that we were interested in understanding CS/IS professors’, students’, and industrial practitioners’ perceived benefits and risks of LLM use as well as ethical concerns in educational contexts. Potential participants were invited to an interview that took no more than 1 hour. Some non-CS/IS majors (e.g., U1 and D2) who had extensive experience in coding and using LLMs also volunteered to participate. We stopped the recruiting process after reaching planned numbers, which also felt like the point of saturation. The participants' demographic information can be seen in Table~\ref{demographic} in the Appendix. All participants were located in the US and people identifying as male or female were equally represented.

\subsection{Interview Process}
After introducing the research project and obtaining participants' verbal consent for interviewing and recording, the first questions asked were about their experiences and perceptions of LLM use in their learning, teaching, or jobs. We asked about their concrete use cases and the perceived role of LLMs in the human-AI collaboration process. For example, students were asked if and how they used LLMs to help with their assignments, essays, course material understanding, etc. Then we went in-depth, asking about the participants' ethical concerns and perceived risks of LLM use in educational or professional settings. The ethical discussion was driven by five prepared topics informed by our literature review, namely, inaccurate answers, hallucinations, bias, privacy, and academic integrity. 
% and human autonomy. 
For instance, during the discussion of privacy risks, we asked, ``Have you previously worried about the risk of privacy leakage when interacting with LLMs? Have you experienced any privacy leakage? How do you think it will affect students and professors/users?'' The detailed interview script is attached in the appendix (Section \ref{script}). The lead author conducted all the interviews in English on Zoom and recordings were then transcribed. The study was IRB-approved.

\subsection{Data Analysis}
We used thematic analysis \cite{braun2012thematic} for the qualitative data analysis. The lead author conducted open inductive coding on the detailed notes taken during the interviews and the transcripts, and discussed the results with the team. Different levels of themes and quotes were organized into a hierarchical structure using XMind, a mind-mapping tool, on which the research team reached a consensus. Below, we use pseudonymized, lightly edited quotes to illustrate our findings.

%% file: 4-results.tex
%%%%%%%%%%%%%%%%%%%%%%%%%%%%%%%%%%%%%%%%%%%%%%%%%%%%%%%%%%%%%%%%%%
\section{Results}

Three main themes emerged during the analysis process, including ``ChatGPT use and experience,'' ``ethical concerns,'' and ``education policy.'' Under ``ChatGPT use and experience,'' several subthemes emerged such as ``overall experience and perception,'' ``ChatGPT's role,'' ``comparison with other AI tools,'' ``specific uses,'' and ``prompting skills.'' Ethical concerns discussed by the participants were more predictable given the structured nature of this part of the interview questions; nevertheless, the participants expressed several additional concerns. Under the theme of ``education policy,'' subthemes such as ``don't ban ChatGPT,'' ``restriction of ChatGPT use,'' ``rethinking education,'' ``proactive education of digital literacy,'' and ``lack of accountability'' emerged. 

\subsection{ChatGPT Use and Experience}
Our participants used LLMs, mostly ChatGPT-3.5, for writing (N=16), Q\&A (N=15), and coding (N=14) in diverse academic and industrial settings. Other LLMs, like those powering Baidu' Ernie or Microsoft's Copilot, were mostly mentioned as comparisons. This indicates that interviewees understand the utility and ethics of LLMs through the lens of their experience with ChatGPT. At the same time, some interviewees were very well-versed in LLMs, having researched them, finetuned their own models, etc.

% three industrial professionals and 8 out of 12 students at different levels.

% They saw ChatGPT as a tool (N=7), an assistant (N=6), a tutor (N=5), but seldom a peer to chat with (N=2). 
% Although LLMs can generate on-topic assessment tasks for instructors \cite{onal2023cross}, the professors interviewed did not extensively use ChatGPT for teaching due to a lack of necessity or skepticism of its value. 

Most participants perceived ChatGPT positively as ``the great invention of humans'' (U2), an efficiency booster, and a good tool for education. Compared to other AI tools and expert systems that our participants had used, ChatGPT had multiple advantages. For U3, ChatGPT users enjoyed more \textbf{autonomy and control} compared to other AI tools: \textit{``It's an active use of AI. You’re taking control. Other products provide AI-generated things for you, and you can use them in no other way.''} I1 compared the \textbf{general-purpose} ChatGPT to other specific-purpose AI tools such as TikTok filters, \textit{``[Other AI tools are for] specific usage and can’t be asked questions. ChatGPT is a universal, general agent. You can ask ChatGPT everything.''} I3 and I4 valued the \textbf{personalization} ChatGPT affords compared to other AI implementations.

% Three participants talked about the \textbf{personalized experience} when interacting with ChatGPT, in terms of both user input and ChatGPT responses (I4). I3 further coined a term, ``individualized personalization,'' for the individualized experience of learning with and from ChatGPT. Traditional AI, on the contrary, only ``bucketed users'' and did not provide a high level of personalization. 
% They valued LLM capabilities for being general-purpose, providing personalized experience, creativity, facilitating free-form interaction, providing autonomy and control for users, and being inclusive to novice users.

Disadvantages of ChatGPT included token limitation (M1, I4: ``Claude can handle longer text''), generating hallucinations (P3), deep fake risks (U4), lack of integration with search engines (D1: ``Copilot is better than ChatGPT in coding as it's combined with search engines''), and being close-sourced (D4: ``CLIP is trainable and fine-tuning is affordable'').
Participants typically regarded ChatGPT outputs with a grain of salt and applied their own critical thinking to learn, to make sure the programs were correct, or to prevent essays from being identified as ChatGPT-generated.

Participants valued prompting skills to make the best use of ChatGPT's powerful Q\&A and feedback capability \cite{jacobsen2023promises}, echoing previous research \cite{jacobsen2023promises, rajala2023call}. 
% Prompting skills were important to realize the full potential of LLMs. 
Several participants (N=5) adopted a ``multi-round and drop'' approach when prompting ChatGPT. U1 was a typical case, \textit{``I asked the question two to three times. If the answer is still unsatisfactory, I switch to Google.''} M2 would enter more details in the prompt to see if ChatGPT generated better responses. P3 saw this as a trial-and-error process. D3 suggested prompting ChatGPT in a clear structure, such as what content she wanted to express in a paragraph. M4 felt frustrated when prompting ChatGPT, as it only gave him bullet points. He acknowledged that he was not a good prompt writer and would give up after a few rounds of conversations.

\subsection{Ethical Concerns}

\subsubsection{Inaccuracy}
Twelve participants recalled inaccurate responses or factual errors when interacting with ChatGPT. Repeated incorrect answers upset U3. D3 estimated an accuracy rate of 60\%-70\%, so she always needed to use her judgment. D1 similarly gave an accuracy rate of 60\%, but both of them acknowledged that ChatGPT was more accurate than search engines (30\%-40\% accuracy, D1). P2 attributed inaccurate answers to the fact that ChatGPT always generated an answer with confidence and never said ``I don't know.'' U4 echoed this opinion, \textit{``It writes confidently even if it’s wrong''}. When D1 tested ChatGPT on history knowledge, such as historical Chinese people, it often erred.

ChatGPT may generate code with bugs or misunderstand user instructions, as in I1's case, so he had to manually check ChatGPT-generated code. He noted that incorrect Linux commands provided by ChatGPT could be dangerous, destroying people's previous work. 
His countermeasure was to understand each line of the code. I4 similarly expressed the quality of ChatGPT-generated code was sometimes poor. 
U2 did not find ChatGPT believable anymore after it suggested non-existent elements in an API.

The inaccuracy of ChatGPT responses may naturally lead to students' low grades in assignments and essays (U4). M2 thought it was the student's responsibility to identify wrong answers. She would ask ChatGPT 5-6 times until it gave the correct answer and showed more details. On the contrary, I4 thought students were prone to get misleading answers since they lacked specific training in the domain.
% TAs knew much better than ChatGPT about domain knowledge and thus should label and enhance the dataset ChatGPT was trained on. 
P2 thought professors should communicate with students to help them understand this issue and make good use of ChatGPT. 
% \textit{``If students turned in the wrong ChatGPT-generated responses at the last minute, they are going to lose points.''}

\subsubsection{Hallucination}

Hallucinations are not an ethical concern per se, but they do implicate the reliability and trustworthiness of LLMs. Hallucination was prevalent in our participants' LLM use, especially when ChatGPT responded to domain-specific questions. 

Five participants discussed hallucination in code or algorithm generation, e.g., suggesting non-existent functions and libraries (I1), making up HTML elements (I3), and using non-existent figure plotting packages (D3), etc. 
% P2 used ChatGPT to construct an AI algorithm and the algorithm tended to hallucinate output. 
% D2 attributed hallucination to the fact that ChatGPT was a generative model rather than an extractive model.
Eleven participants complained about hallucinated references from ChatGPT, which hurts its utility as a literature review tool. 
% (D2) and made people feel uncertain (M4).
% P1 thought ``mixing things up'' was strange since the papers were all in Google Scholar. 
P3 suggested other customized LLMs that give better references. 
% U2 similarly mentioned experiences of getting irrelevant summaries of papers and suggested using other customized tools. 
% M1 always double-checked ChatGPT-generated references. 
% P2 thought the fake references were still helpful because they helped him search and find real papers by himself by providing him with useful keywords in the titles. 
% \textit{``If you understand its mechanism, it’s fine.''} 
% P3 provided an interesting perspective, suggesting that students would never write their papers again if ChatGPT never hallucinated and wrote well. 
% P1 was worried LLMs threatened the integrity of scholarly works, with a lot of research not supported by anything and no system to deal with this issue. 
P1 suggested that ChatGPT may help in checking references, instead of worsening the case, if used wisely.
% She was inclined to rely on students to use sources correctly. 
P2 took an ontological approach, teaching students to think about what is truth, which ``became a more important digital literacy skill in the era of ChatGPT.''

\subsubsection{Bias}
Nine participants were not worried about bias carried by ChatGPT. U2 observed that ChatGPT was good at avoiding answering potentially biased questions. Often, it gave a disclaimer, \textit{``I’m just a model...''}
Most participants felt bias in LLMs may impose stereotypes or Western-centric views, but contextualized the risk.
% I2, on the other hand, thought such responses were not biased, whereas, were also not unbiased. 
% She thought students in the US could easily identify biased responses, whereas, in more traditional cultures, people were more likely to be affected. 
I1 thought university students could understand and recognize bias but it was more of a concern for younger students. M4 acknowledged ChatGPT's bias but thought it displayed bias less than search engines and many of his professors. 
He liked the safety feature of ChatGPT where people could report harmful responses. 
Similarly, U3 compared ChatGPT to Baidu’s LLM product, Ernie, which gave explicitly biased answers. She thought ChatGPT already filters most real world bias.

Some participants proposed countermeasures (N=5) to mitigate the impact of biased responses. M3 tried to prevent bias by prompting ChatGPT appropriately. P1 actively taught about bias in her class, for example, asking students to discuss gendered language, or what a computer programmer looks like. P3 suggested not using AI in critical applications: \textit{``The bias in AI is based on how the world runs. [...]. Maybe just don’t use it in certain systems such as hiring.''} 

\subsubsection{Privacy}
People had varied levels of concern regarding privacy leakage when interacting with ChatGPT, and four participants were not concerned at all. Some participants viewed the privacy issue contextually. P1 thought the privacy issue was not as severe as other LLM ethics issues: \textit{``Google has tons of user information. We don’t know how it’s used, but we still want to use free Gmail.''} I4 criticized OpenAI for transparency, stating that dark patterns tricked people into giving their data for training ChatGPT: \textit{``It's not legal to use your data for training, but they make it look legal.''} U3 thought ChatGPT had advantages over other AI-powered techs regarding ethical concerns: \textit{``[ChatGPT] does its best to protect user privacy and reduce bias. OpenAI hasn’t sold my data to anyone yet. I haven't heard of any data leakage scandals. There are much more problems with other technologies, like scams.''} Four participants claimed they were unconcerned with privacy but would still be cautious in what they wrote to ChatGPT.

\textbf{Privacy risks and mitigations.}
Five participants expressed explicit concerns regarding privacy leakage. Such a mindset translated into privacy-preserving practices in their interaction with ChatGPT, e.g., not providing personally identifiable information (D1), and keeping questions as vague as possible (U4). Some students were concerned that OpenAI would report their ChatGPT usage to professors. Thus, privacy can interact with academic integrity. Even without a privacy/security leak, students generally thought ``professors will know'' if they tried to cheat. 

\textbf{Participants' concern for others' privacy precluded certain LLM uses.}
Eight participants expressed a higher level of concern over feeding their students', clients', or research participants' data into ChatGPT than their own data, e.g., \textit{``I can’t speak for them (M3),''} \textit{``I respect and cannot manipulate student concerns since there is a power balance (D2),''} and \textit{``Not supposed to share students' grades with anybody (P2).''}  M4 elaborated, \textit{``I don’t give participants’ or clients’ information to ChatGPT. The contract is signed. It’s more about protecting other people’s privacy risks.''} 
% I3 did not give out confidential company information and customers’ account information.
% She also did not give her ``really personal information'' like social security number, address, and kids’ identity information in the conversation. 

\textbf{Privacy and equity.}
Several interviewees proposed that LLM privacy is not equitable and furthers a digital divide.
U3 pointed out that the paid version of ChatGPT had better data protection over user data, treating privacy as a privilege. U2's university provides ChatGPT premium, so has the option to use it. Still, U2 is conscious of privacy problems while using the university version of ChatGPT or using ChatGPT with university credentials.
 
\subsubsection{Academic Integrity}
Nine participants raised cheating as a concern for academic integrity, because it may hurt students' learning efforts. U3 acknowledged: \textit{``Sometimes I hit ChatGPT with all the questions without thinking by myself. I do this even if I have AI literacy. I'm not learning from the problem-solving process.''} A lack of shared understanding of what constituted cheating, knowledge, or education further complicated the issue. I1 illustrated this point, \textit{``How do you define you’re studying? What is knowledge? Before, we learned and memorized knowledge from previous people. Now we just ask ChatGPT. I don't know if students learn less, but more efficiently for sure.''} 

% M4 provided a hilarious analogy to describe cheating students -- \textit{``Student forums become Reddit forums of chatbots.''} 
% P3 thought cheating in HCI courses was less of a concern -- students had to do usability testing or generate design prototypes, which could not be done by ChatGPT, whereas assignments were only worth a small portion of the grade. 
% She commented on cheating in coding-oriented courses, \textit{``Even if you got A's in all courses, your interview is what matters for job seeking.''}

\textbf{Use or abuse?} Six participants believed cheating only occurred when ChatGPT was misused; otherwise, it was a good learning tool. P1 noted, \textit{``I allow students to use it, but ask them to disclose prompts used. It’s just a tool. Using tools is not cheating. People used to hand-write at some point. Later, we use computers to type. Prohibiting it [ChatGPT] is a waste of time. We need to prepare students for the workplace. It's not realistic to ask them to not use.''} D4 thought using ChatGPT to generate the whole essay constituted plagiarism while using it to refine writing was acceptable.
% \textit{``Research problems are specific. You should be the one to write if you’re an expert. ChatGPT is not domain-specific and only gives generic responses.''} 
% D1's redline was to use ChatGPT without people's own ideas.

\textbf{Detection.} U4 suggested using detection tools to curb ChatGPT's strong generative power in helping students cheat in exams. However, three other participants mentioned potential errors and biases introduced by detection tools. D2 refused to use a tool to check students’ work. She thought it was a big issue if false positives happened, which was going to be a disaster for students. Alternatively, she would ask students to talk about their thought processes for writing and research to learn from the process. 
% I3 believed there was no good way to detect ChatGPT-generated essays. 
U1 argued that translated content was more likely to be classified as AI-generated, leading to more false positives for non-native English speakers who used translation tools for their assignments.

\subsubsection{Other Ethics Issues}
Several participants noted LLMs improve the accessibility of education. M4 thought ChatGPT facilitated natural interaction and removed the need to interact with APIs: \textit{``People can easily decipher the responses from ChatGPT. It's good for non-tech-savvy users and enables everyone to use AI.''} U2, an EdTech researcher and equitable education promoter explained how ChatGPT disproportionately helps neuroatypical students: \textit{``Students with mental issues or difficulties need ChatGPT more. Previously they were like, `I couldn’t do it,' and late dropped the course. Now they are like, `I’ll manage to do it [with ChatGPT].' You can’t say they’re using ChatGPT unethically.''} 

Several participants noted an unclear line between LLM use and over-reliance. For example, M1 reflected on how over-reliance on ChatGPT took away opportunities for learning, \textit{``At some point, I’m over-relying on AI, for example, when I’m coding. I miss the chance to explore how to code and learn from errors and debugging. It’s just too efficient.''} 

% \zakcomment{Concerns about LLMs}
% Simultaneously, they expressed various concerns over robustness, human autonomy, inaccuracy, and so on. 

% \zakcomment{Concerns about consistency of LLM performance}
% U4 elaborated on this aspect and mentioned its occasional breakdowns: \textit{``I have a good perception of it overall [but] it wasn’t working as well in a certain month. The responses didn't make sense. It summarized something but it was completely different [from the original content]. It stopped working sometimes and showed Internet errors.''}

%%%%%%%%%%%%%%%%%%%%%%%%%%%%%%%%%%%%%%%%%%%%%%%%%%%%%%%%%
\subsection{LLM Usage and Ethics in CS Education: User Mental Models}
To unite LLM usage patterns with ethics, it can be useful to think about user mental models. As discussed, previous works have called for work to better understand the mental models of LLMs in CS education \cite{lau2023ban}. 
% According to a systematic literature review, 
Mental models have appeared in HCI for decades, but understandings greatly vary about what a mental model entails. 
Researchers often do not define mental models or use very vague definitions, like ``internal representations of the external world by humans'' \cite{hu2023scoping}.
We adopt the following definition of mental model, which is more suitable for user studies, and is present in design and usability literature -- mental models are ``the subjective and hypothetical mental representation that integrates an individual’s memory, knowledge, perception, and assumptions of the target system'' \cite{hu2023scoping}. 
% \zakcomment{(Hu and Twidale, 2023:114)}

The authors discussed and co-designed this representation of mental models iteratively based on participants' responses (i.e., transcripts, qualitative analysis thereof).
We started from ``main use cases'' which are broad goal contexts wherein the majority of interviewees use LLMs.
Other mental models were considered but excluded based on how many interviewees explicitly indicated them.
For example, two of four teachers discussed using ChatGPT as a grading tool, but this is only 10\% of our sample, so we do not feel we have enough evidence on this aspect.
The models changed over time. For example, rather than ``Information Tool'', the third mental model was initially Q\&A. After discussing we agreed that Q\&A is part of a larger task context, which relates to retrieving and summarizing information. We collectively reviewed and discussed evidence to further distill and refine the model to be concise and informative.

\begin{figure*}
  \includegraphics[width=0.8\textwidth]{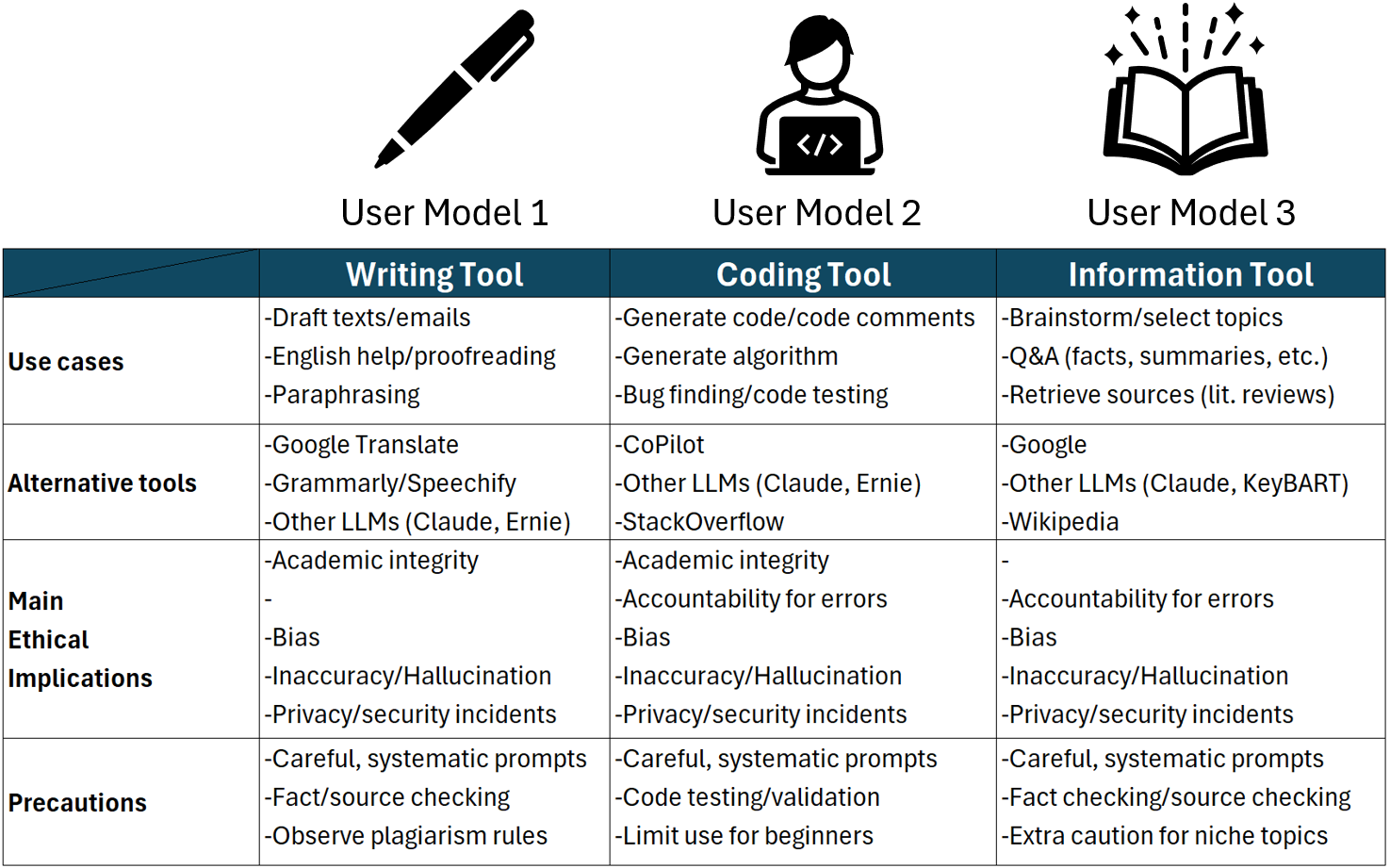}
  \caption{LLM user mental models in CS education.}
  \label{mental:model}
\end{figure*}

We propose that LLM users in CS education primarily have three mental models for LLMs: Writing Tool, Coding Tool, and Information Tool. 
Each model has specific use cases, possible alternatives (e.g., CoPilot for coding), main ethical implications, and practical precautions, as shown in Figure \ref{mental:model}.
The model sacrifices comprehensiveness for comprehensibility.
We do not present the model as a totalizing encapsulation, but as a useful simplification to guide education policy/governance and future research.
For example, there is a crossover between Writing Tool and Information Tool, but we feel that the evidence points to different practical and ethical considerations, justifying separate categories.

\textbf{Mental model 1: LLMs as Writing Tool.}
Most interviewees (16/20) used LLMs as a writing tool to write new texts or improve existing ones.
While most students claimed not to use LLMs to write essays/assignments per se, LLMs still helped write emails, draft ``first try'' paragraphs in academic essays, and paraphrase prompts into academic style. 
For the most part, prior to LLMs being deployed in convenient chatbots, tools could translate and grammar/spell check, but no comparable tools existed that could draft unique text. 
A few interviewees who were not English-native speakers used LLMs more heavily for proofreading and also for translation tasks.
Participants generally expressed that as a writing tool, LLMs pose the greatest risk for academic integrity, requiring very precise plagiarism definitions (which are typically lacking).

\textbf{Mental model 2: LLMs as Coding Tool.}
% Most student participants did not use ChatGPT for writing-based assignments, mainly due to professors' rules or caution against ChatGPT use and willingness to learn.
Most interviewees (14/20) used LLMs to assist them in coding tasks. 
Here academic integrity was less of an issue, as code plagiarism has different standards than writing (e.g., copying a function from StackOverflow is not usually problematic) \cite{lakhani2003hackers}.
However, participants emphasized the possible tradeoff between ``doing coding'' and ``learning coding'' for beginners.
Additionally, many emphasized the responsibility of coders to understand and test code, especially before deploying it in sensitive contexts.

\textbf{Mental model 3: LLMs as Information Tool.}
Our participants (15/20) also used LLMs for information retrieval and summarization.
Many participants discussed the advantages and disadvantages of ChatGPT for this usage, and shared when LLMs work better for them than Wikipedia, Google, and other tools.
Here, the novelty of LLMs is less considerable than writing or coding, but ChatGPT was a useful alternative to search engines or asking a human.
For example, treating ChatGPT as a Q\&A bot was very helpful in quickly summarizing an unfamiliar topic, retrieving a good definition, etc. 
A few participants also used LLMs where they are likely to fail, such as accurate sourcing of information for references or literature review.

%%%%%%%%%%%%%%%%%%%%%%%%%%%%%%%%%%%%%%%%%%%%%%%%%%%%%%%%%%%%%%%%%%%%%%%%%%%%%
\subsection{Policy and Governance of LLMs in CS Education}

% \subsection{Education Contexts}

\subsubsection{Don't Ban ChatGPT, Regulate It} All participants agreed that ChatGPT should not be banned, as it was useful for learning and was prevalent in the workplace. Universities should prepare students for the LLM era and in the meantime adopt necessary regulations and restrictions to help them learn. U1 gave an example of how ChatGPT helped students learn to code, \textit{``Professors should not ban AI uses. AI is quite useful because it can generate code correctly. It's useful for students to learn to code. Professors may not have enough time to teach it in class.''} 
% At the same time, he was not against restrictions on AI use and suggested 30\% as a threshold for the reuse rate of ChatGPT-generated code. 
% I3 noted, \textit{``Don’t ban it. Integrate it into class and let them learn its limitations and benefits.''} D2 also suggested administration not be overly cautious about LLMs or prohibit their use.

Banning ChatGPT was not practical, as professors could not check every student's work closely. U3 said, \textit{``It's not possible to ask students to not use it. Teachers can’t check everything. It's not possible to ban ChatGPT.''} P2 agreed and urged educators to think about students' role in the collaborative learning process, \textit{``Banning it [ChatGPT] is a bad idea and not practical. See it as a collaborator or tool that helps students learn.''} 
% I don't know its effect on our brains yet. We need to think about how students learn with the tools. Do they have a role there? 

Another consensus was that students should not use ChatGPT for everything and restrictions should be in place. M3 leaned toward more restricted LLM policies given the ``lazy nature of human beings,'' \textit{``We lean toward doing low-effort, high-reward things as humans. Students are still growing. Their lives will be destroyed if they are not instructed well. Over-reliance on ChatGPT to skip work is not good for human development.''} M2 mentioned an incident in a co-op course for preparing students for internships and full-time jobs, where students used ChatGPT to write their cover letters and emails intended for HRs even after the teacher had warned against such use. 
% The teacher became angry after finding similar expressions in students' emails, saying, \textit{``If HR notices it, it will cause you to lose jobs.''}
M1 thought freshmen and entry-level students should use ChatGPT less in their coursework, otherwise they would lose the chance to learn to code. 
% Graduate students, on the other hand, already had coding skills and the copyright issue was more of a concern when writing papers; software engineers were also able to make good use of ChatGPT because they had skills in CS and models.
% What our participants wanted to preserve the most were learning outcomes (N=5), creativity and critical thinking (N=2), and academic integrity (N=1).

University policies regarding ChatGPT use should be flexible and contextual, according to our participants. D4 believed ChatGPT use in different scenarios needed to be treated differently -- using ChatGPT to answer homework questions was not acceptable while using it to refine reports was fine, as long as the students highlighted the parts generated by ChatGPT. P2 similarly emphasized that deciding ethicality of ChatGPT use was based on context, stating that policy had to be nimble and not once and for all. M4 was against a university-wide policy given the major differences between subjects, and instead for department-level policies.
% through the collaboration between the department director and students. 
Some participants did not think university policy should exist regarding ChatGPT use before AI detection tools were in place (D1). P1 further added that ChatGPT was very new and more observations were needed to inform an effective policy, \textit{``The tool is too new. We need more data to understand how students use it. Having a policy now won’t address what’s happening. I'm very wary and would oppose a policy. We need to think over time. It's like book banning.''} P4 echoed this opinion, saying universities should not step in too early.

\textbf{Lack of clarity in accountability.}
% \zakcomment{I think we cut most of this section and move the good sentences elsewhere.}
There was not a consensus regarding who was responsible for students' ethical ChatGPT use. Some thought students held responsibility. For example, P1 said students were responsible as it was their learning process and they decided what was integrity. 
% U2 echoed this view, \textit{``We are the ones who use ChatGPT. We bear the risks of over-reliance on ChatGPT and not getting anything at all, for example, performing poorly in later job interviews.''} She thought it was hard for professors to regulate ChatGPT usage by students, as ``students will always find a way to use it.'' 
% M2 thought professors were responsible since each class was different. 
% U2 was not sure to what extent instructors could intervene given the volume of student work they had to process.
Five participants put the responsibility on universities or departments. I1 thought departments should start from the high level to guide teachers and students since students and teachers were confused, \textit{``The teachers are confused as well. They don’t know where they can take the students. For example, `Do I need to ask the students to memorize the functions, theories, etc.? Or just let them use ChatGPT?' ''} 
% Some suggested more technical actions the university could take. For example, U2 thought students should be connected to internal university databases to prevent hallucinations; U4 suggested universities create new LLMs to help students learn as ``the current ChatGPT was too powerful.''

Five participants agreed that it was the joint responsibility of students, professors, and universities to make students ethically use ChatGPT. I3 elaborated on a tiered system engaging students' honesty, professors' rules, and universities' judgment if the case was escalated, \textit{``In an ideal world, it's students' responsibility. But see all those cheating. Everyone is responsible. Teachers need to enforce rules. If they are not followed, there is a need to  escalate it to the university level.''} U3 described a similar responsibility distribution.

\textbf{Rethinking Education.}
In the ChatGPT era, we need to rethink education, assessment, and academic integrity.
Re-constructing assessments and assignments was commonly expressed as a solution. I1 cited a quote by Andrew Ng, \textit{``If your assignments can be done by ChatGPT, you’re not a good educator. We need to find a way to prepare assignments that ChatGPT cannot do.''} 
I2 used to be a teaching assistant and tended to ask students to write essays that could not be generated by ChatGPT. She would ask students to write specific points, e.g., how a particular argument connected back to a particular lecture. P1 elaborated on the vague nature of the current writing-based assessment, \textit{``We need to take a scaffolding approach, telling students what the goal is for the assessment. For example, what is understanding? We need to think of other ways to demonstrate understanding than writing.''} I4 similarly suggested designing assignments more carefully, whose answers should not be openly accessible on the open Internet. 

I1 suggested departments think about what knowledge is, redefine study, and make policies accordingly based on their high-level understanding. According to him, \textit{``What used to be knowledge may not be knowledge anymore, like the case of Google search.''} U2 would like to see changes to the syllabus, with a new focus, and including more authentic content to prepare students for a future where AI is everywhere. 
% She noted, \textit{``It should be more than memorizing the facts.''} 
U4 thought the curriculum needed to be updated and made more challenging. 
% He further suggested having in-class instead of take-home essays. 
With new definitions of knowledge, D2 wanted to see developed, constructive policies and wording regarding what constituted plagiarism in this new context.

\subsubsection{Proactive ChatGPT Education for Both Students and Professors}
Instead of regulating ChatGPT use, more participants wanted to see proactive education to teach students how to use it ethically and responsibly. P3 thought tech ethics was always a conversation and should be integrated in Day 1. She told us how she would convey the message to students in her course, \textit{``In my class, I'll tell my students, `You’re gonna use it [ChatGPT]. ChatGPT is a baseline and you need to make your answers better than that.'''} 
% I3 thought educating students to use ChatGPT properly was easier than prohibiting them from using it. 
From a student's perspective, M4 suggested having courses on how to use ChatGPT. He thought how to use ChatGPT and eliminating its misuse were important parts of digital literacy in the LLM era. Though P3 had not seen university rules or professors prohibiting ChatGPT, she thought universities should take the responsibility of delivering ethics courses to teach and educate students to make their own decisions. 

% Similarly, U3 encouraged teaching students how to better use AI and increase their AI literacy. 
% She thought ChatGPT was not drastically different from search engines or books, but just gave answers in a different way, and thus should not be ``demonized.''

Professors should be educated about ChatGPT in a way that allows them to realize its benefits for student learning, as D3 explained, \textit{``Efforts are needed to make professors allow ChatGPT use. ChatGPT can help students learn. It might have been the same situation for search engines a few years ago.''} M3 similarly suggested using the search engine analogy to help professors understand ChatGPT's role, \textit{``They should know these tools are not demons to make students lazy. It's mostly about their attitude toward LLMs. I'd compare them [LLMs] to Google and tell them [professors] it’s impossible to prohibit it [LLM use]. It's more about how to use it instead of whether to use it.''} The professors also discussed the possibilities and benefits of workshops for themselves. For example, P3 suggested education for professors to learn how to use ChatGPT in teaching and research, \textit{``You don’t know how to make it benefit you unless you play with it. I'd like to see workshops for faculty on how to effectively use ChatGPT in teaching or research.''}

%% file: 5-discussion.tex
%%%%%%%%%%%%%%%%%%%%%%%%%%%%%%%%%%%%%%%%%%%%%%%%%%%%%%%%%%%%%%%%%%%%
\section{Discussion}

% Aligned with the prior literature on LLMs in higher education \cite{laato2023ai, ansari2023mapping, aithal2023application, perera2023ai, wang2023navigating, vargas2023challenges, fuchs2023exploring, tajik2023comprehensive, dempere2023impact}, our participants dominantly used ChatGPT over other LLMs.

% \subsection{How CS Students and Professionals Use ChatGPT}

% \begin{tcolorbox}[width=\linewidth, colback=white!95!black, boxrule=0.5pt, left=2pt,right=2pt,top=1pt,bottom=0pt]
% \stepcounter{finding}
% {\bf Finding \arabic{finding}:}
% {CS students, professors, and professionals use LLMs in diverse settings with a grain of salt.}
% \end{tcolorbox}

\subsection{Ethics of LLMs in CS education}

% Even students themselves were aware of student laziness caused by LLMs and were worried about being replaced by the tool in the workplace \cite{yilmaz2023augmented}. 

The views of the participants on ethical concerns raised in the literature confirmed the relevance of hallucination, bias, inaccuracy, etc., for LLMs in CS education.
Additionally, our participants raised diverse concerns, ranging from potential monopoly to energy consumption in training LLMs. 
We also find evidence of the complex interaction of LLM usage patterns and ethical concerns.
For example, our participants commonly opined that novice coders would not learn with ChatGPT's generative power, which could result in deploying unethical code.
This means that the tension between LLMs as a learning tool and LLMs as a coding tool also implicates accountability and responsibility.
Other remaining challenges include inaccurate or biased detection tools \cite{orenstrakh2023detecting, mischak2023challenges}, and a lack of shared understanding of plagiarism, educational assessment, and knowledge formation, which we will elaborate on in the section below. 

Almost unanimously, interviewees believed that critical thinking and verification are required to combat inaccuracy and hallucination -- further, students should be held responsible for LLMs-generated content text or code, echoing previous research \cite{laato2023ai}. 
The influence of anticipated guilt, detection probability, sanction severity, and personal reputational risk on students' ethical use of LLMs \cite{mvondo2023generative} was supported by our sample.
% For example, \textbf{inaccuracy and hallucination} may lead to bad grades and a disrupted learning process, regardless of plagiarism. 
% \textbf{Hallucination} of references hurts scholarly communication; hallucination in essays and assignments makes them more identifiable as AI-generated content; hallucination in code may lead to security misconceptions in students -- one participant deferred the role of checking ChatGPT-generated code to the code review team. 
% There are two approaches to leveraging LLMs in problem-solving, namely, the LLM-first approach and the self-first approach \cite{kumar2023impact}. If students adopt an LLM-first approach, it is important to educate them to apply their own critical thinking. 
% Learning Analytics (LA) may play an important role in understanding and optimizing responsible and reflective use of LLMs in academia, but it was important to first effectively differentiate between AI-assisted and AI-complement actions \cite{hernandez2023chatgpt}. 

Regulating ethical risks embedded in LLMs that do not exclusively pertain to educational or academic contexts, e.g., privacy is important too. 
Our participants are worried about students', research participants', or clients' privacy even if they are desensitized to their personal privacy. 
Professors refuse to feed student assignments into ChatGPT due to concerns about student privacy, copyright, and consent. Students are worried that privacy leakage will lead to the revealing of their ChatGPT use. 
We find support for a privacy gap that could widen with time -- the paid version of ChatGPT affords better privacy protection, threatening equitable educational opportunities and inviting further digital divide \cite{bordt2023chatgpt, lau2023ban}.
Our findings contrast with Shaengchart et al., who did not find concerns of privacy a relevant factor in effectively utilizing LLMs \cite{shaengchart2023factors}. 

% Lastly, most participants are not worried about \textbf{human autonomy}, i.e., being replaced by ChatGPT, especially if their job requires innovation and communication.

% There are distinct concerns that (1) AI will be deployed for an inherently unethical purpose, and (2) AI will be deployed for an ethically appropriate reason but will produce unethical results as a result of incorrect/problematic outputs, e.g., due to technical failures. In the context of LLMs in computing education, (1) might include students using ChatGPT to plagiarize, while (2) might result in a teacher using tools to detect ChatGPT-generated content, but the tool gives false positives. Either type of concern could have a direct and severe impact on student learning.

\subsection{Adjusting Policy and Education}
LLMs have brought two major challenges to CS education -- first, LLMs might impact the effectiveness of student learning; second, LLMs might make assessing students much harder for the teachers. 
However, instead of banning LLMs, which is impractical and likely to fail, our study suggests that universities should prepare students for the LLM era and in the meantime adjust policy and education to help students learn. 

% The evidence suggests several findings. First,
We should think of what could be considered ethical/reasonable use of LLMs in education. One way is to look at earlier technological advances as precedents, like how math curricula adapted to the spread of calculators \cite{waits1997role}. Any education policy on LLMs needs to clarify the boundary between acceptable and unacceptable LLM usage as precisely and specifically as possible. 
Based on our study, acceptable usage presumably involves students acknowledging that they used an LLM/citing it; students shouldn't present an LLM's work as their own. 
Much trickier is defining what uses are acceptable in a CS context, as reusing open-sourced code has become a norm \cite{lakhani2003hackers}. 
% Tricky real-world cases should be considered: for example, if an international student in a US CS program writes a paper in their mother tongue, then uses ChatGPT to translate it, and submits the translation and acknowledges the tool's usage, would this be a violation of policy? Why or why not? 
% If a CS student submits a coding assignment consisting of multiple classes and functions, and one of the functions is exactly what an LLM generated based on a student's own prompt, would this count as plagiarism?
% To the extent possible, any education policy on LLMs should be enforceable, even if we lack tools to identify plagiarized LLM-written code or text. However, any policy must acknowledge that LLMs do create a new way that students can cheat. The policy in CS education needs to consider all three mental models of LLM use: as coding assistant, writing assistant, and information assistant, and define cheating/acceptable usage.
Failing to generate clarity on LLM usage in CS education will have a few impacts. 
Some students who could benefit from using LLMs will not, for fear of violating rules of academic integrity. 
This is an opportunity cost, and it may be especially problematic for international and EFL students. 
Also, well-meaning students who use LLMs risk accidentally running afoul of academic integrity rules. 
This is a direct cost. 
Finally, curricula, especially for entry-level CS courses, may stay poorly suited for assessing student learning outcomes, hurting students' learning and fairness.

% CS instructors have somewhat mixed opinions on LLM use in CS education: some leaned toward banning LLMs and continuing the teaching of programming fundamentals, while the other side wanted to integrate LLMs into courses to prepare students for future jobs \cite{lau2023ban}. CS instructors expressed such concerns as academic fairness (e.g., whether using ChatGPT's free or paid version made a difference); and long-term impact on students’ mental models (e.g., how students trust/accept ChatGPT by default, or check/verify its outputs) \cite{wang2023exploring}. Laato et al. proposed implications such as LLMs' roles in supporting students’ idea formation and critical thinking, possibilities of circumventing copyrights and plagiarism detection software, consequences of over-reliance on LLMs, opportunities for personalized learning, and students being held responsible for LLM output \cite{laato2023ai}. Laato et al. argue that these challenges require instructing students on ethical LLM use.

% ChatGPT can play an important role in education. 
% For students with disabilities or those who are unable to attend traditional classrooms, ChatGPT can improve access to educational materials \cite{rawas2023chatgpt}. One of our participants also stressed ChatGPT's role in helping students with learning difficulties. Broadly speaking, LLMs can provide personalized, on-demand assistance to students. \zakcomment{need to add use for ESL/international students with language challenges.}

\textbf{Sociotechnical governance needs.} LLMs challenge existing contextual norms and are employed outside the scope of current educational privacy and technology regulations. Only a limited set of US states have legislated broadly to constrain personal information flows to AI systems and support data protection, making it necessary to develop new and responsive governance. As results indicate, privacy is only a small part of the potential harms associated with LLMs in education. As such, a comprehensive data protection approach is needed. In California, under the CCPA, rights to access and correct personal information are tied to rights against the use of sensitive characteristics or their proxies in decision-making for protected classes or regulated contexts -- this would be a solid foundation for addressing issues of bias in LLMs. Overall, LLM applications for educational contexts require more supervision and continuous engineering of the systems to ensure alignment with key values and laws. This would also allow for conformity to unique regulatory regimes globally -- our results highlight the needs of international students studying in the US to comply with considerations, such as GDPR.

\textbf{What do ideal education policies look like?} Fowler et al. pointed out the lack of official recommendations for LLMs' implementation in higher education and stressed the importance of involving all stakeholders in the decision-making process to ensure the successful adoption of LLMs \cite{fowler2023first}. 
% Even ChatGPT itself emphasized the urgent need for clear policies, guidelines, and frameworks to responsibly integrate ChatGPT in higher education \cite{michel2023challenges}. 
% Most existing guidelines were not from universities but from researchers. For example, Gimpel et al. proposed comprehensive lists of guidelines and recommendations for students and lectures \cite{gimpel2023unlocking}. Their recommendations for students included respecting the law and examination regulations, reflecting on learning goals, using ChatGPT as a writing/learning/coding partner, and bewaring of risks when using ChatGPT; recommendations for lectures touched on both the teaching aspect (e.g., reflecting on learning objectives, creating learning materials, and encouraging students to use ChatGPT) and the assessment aspect (e.g., requiring students to declare how ChatGPT and other advanced tools were used, innovating assessment format, and rethinking the evaluation criteria for assignments). 
Rudolph et al. advised against a policing approach that focused on discovering academic misconduct, such as detecting the use of ChatGPT and other AI tools, but favored an approach that built trusting relationships with students in a student-centric pedagogy \cite{rudolph2023chatgpt}. 
% assessments should be \textit{for} and \textit{as} learning rather than solely \textit{of} learning \cite{rudolph2023chatgpt}. 
Our study supports this past work by cautioning against ``once-and-for-all'' policies and instead stressing student learning, and contextual and flexible policy depending on different uses and different disciplines.
% regarding ChatGPT use for several reasons. First, LLMs are a new invention, thus professors and university administrators have not gained a sufficient understanding of how students use them and their potential benefits and risks yet. Second, given the drastic difference between disciplines, a university-wide policy can be hardly relevant; they instead prefer department-level policy or instructors' discretion. 

\textbf{Who is responsible?}
A concerning study finding is that there is currently no consensus on who is responsible for regulating and educating students on ethical LLM use in higher education. Every party seems relevant to the mission -- students are the learners and should be responsible for themselves; teachers need to enforce rules; universities should deliver education on ethical LLM use for both students and teachers; departments should standardize rules between classes, etc. 
More conversations with different stakeholders should be facilitated to delineate responsibility, enable collaboration, and craft suitable governance policies.

\textbf{Rethinking education.} LLMs have the potential to transform the interaction models among learners and teachers \cite{khaddage2023towards}. Dai et al. insightfully conceptualized ChatGPT as a ``student-driven innovation,'' which has the potential to empower students and enhance their educational experiences \cite{dai2023reconceptualizing}. At the same time, they also expressed the need to address emerging challenges regarding student training, higher education curricula and assessment, and technology development and governance. Our participants all agreed that the ultimate goal of education was to enable students to learn. Rethinking education, assessment, and skills is necessary when a new technology is invented and used for learning.
% -- back in 2010, even Web2 was a concerning source of academic misconduct in higher education \cite{waycott2010implications}.
% ; whereas, nowadays, as long as students do not directly copy content from search engines, they will not be punished.
Existing assessment tools such as knowledge-based exams and problem-solving questions may no longer be adequate to confirm students’ learning and performance since ChatGPT does well in these tasks 
% and passed the academic integrity tests by Turnitin 
\cite{chaudhry2023time}. 
% Assessment methods thus need to be updated.
Assignments should be designed in a way that is not doable by ChatGPT alone, e.g., being open-ended, or being relevant to the course materials, as suggested by our participants and raised in \cite{lee2023chatgpt, sullivan2023chatgpt}. 
% The simple heuristic, based on Sullivan et al.'s analysis of news articles \cite{sullivan2023chatgpt}, is to redesign assessment tasks in such a way that they can not be completed as easily by AI tools.
Students should also be trained to do innovative work that can not be replaced by LLMs. Higher education should continue to emphasize the ``soft skills'' required in industry and the value of a ``human touch'' in projects; when it comes to CS education, students’ understanding of security and secure software development practices should be emphasized \cite{shardlow2023chatgpt}.

Lastly, ethical LLM use has become an integral part of \textbf{digital literacy}. Training for educators and students is deemed important by university students to effectively utilize LLMs \cite{shaengchart2023factors}. 
% One participant argued that ethical discussions should happen early, not only in academics but also in any conversation about technology. 
A course to teach students how to utilize LLMs responsibly in learning was favored by both students and professors in our study. 
% Professors also raised the need for workshops to educate them on using LLMs for teaching and research. Students thought educating professors on the benefits of ChatGPT could effectively prevent them from ``demonizing'' or putting overly strict rules over ChatGPT. 
Our findings echo existing calls for training to promote students' digital literacy, including limitations of AI systems and the ability to validate information with multiple sources \cite{barbas2023importance}.

% \textbf{LLMs in computing education vs for-profit IT industry}
% Fewer regulations are in the industry than in education
% The use of LLMs is ethically fraught in the IT industry. For instance, The New York Times sued OpenAI for not paying them to train ChatGPT on their data, i.e., copyright infringement \cite{nyt}. However, as our participants suggested, less regulation will be enforced in the industry given its for-profit nature and practitioners' awareness of the ethical use of LLMs -- which is in question. New challenges may arise in the IT industry. For example, certificates of originality (COO) or their equivalents have been used by IT companies and developers to claim the originality of their code or programs. If and how COO is effective when LLMs are used to help with coding is currently unknown and worth exploring. Questions remain as to how to align education policies regarding LLM use with industrial goals, thus better preparing students for their future careers. 

\subsection{Limitations and Future Work}
Our current study has several limitations. First, we focus on CS-related disciplines as a unit, but in fact this is a heterogeneous group. Future research could compare results for focus groups with professors from different fields/disciplines. 
% It would be interesting to see how the more limited LLM understanding in non-CS settings impacts stakeholder opinions on digital literacy education or ethics.
Second, our sample size is relatively small (N=20) with only 4 participants per stakeholder category, with limited generalizability. For example, the professors we interviewed did not much use LLMs, but this may be unusual.
A survey design would offer more conclusive evidence on broader trends. 
% Third, our observations and inquiries are based in a single country. Comparing LLM policies in the higher education systems of different countries may reveal interesting differences.
Third, while we intended to study LLM attitudes, ChatGPT was the focal point for most available research and interviewee experiences. Still, we feel that our work on the practical and ethical consequences of LLM in CS education would be generalizable to any convenient LLM-powered chatbot. As more LLM products become popularly used, research on LLMs will expand beyond current product offerings.

%% file: 6-conclusion.tex
%%%%%%%%%%%%%%%%%%%%%%%%%%%%%%%%%%%%%%%
\section{Conclusion}
Our research uncovered that CS students, professors, and industrial practitioners have an overall positive experience and perception of ChatGPT, but with a pinch of salt. LLM users' mental models can be categorized into coding, writing, and information tools, fraught with a wide range of ethical concerns.
% People praised it for its general-purpose, personalized, creative, free-form, human-empowering, and inclusive features. ChatGPT served diverse use cases in higher education, such as Q\&A, {writing code, debugging code, generating code comments}, assisting assignments, writing essays, preparing teaching materials, and helping non-native English speakers. 
% ChatGPT was treated as a tool, a tutor, or a peer. 
The multiple stakeholders reached some consensus on education policies: (1) Multiple ethical concerns regarding LLM use directly impact CS students; (2) Contextual and flexible regulations should be adopted to help student learning; (3) Rethink education, knowledge, and assessment, and (4) Provide proactive education for both students and professors to prepare them for responsible LLM use. 

% \begin{figure}[h]
%   \centering
%   \includegraphics[width=\linewidth]{sample-franklin}
%   \caption{1907 Franklin Model D roadster. Photograph by Harris \&
%     Ewing, Inc. [Public domain], via Wikimedia
%     Commons. (\url{https://goo.gl/VLCRBB}).}
%   \Description{A woman and a girl in white dresses sit in an open car.}
% \end{figure}

%% file: 8-interview.tex
\newpage
\section{Interview Script}
\label{script}
Note: this has been lightly edited for length and privacy.

\subsection{Introduction}
[REDACTED]

\subsection{Interview Questions}
\noindent \textbf{General Experience of LLM Use}
\begin{itemize}
    \item Could you tell us a little bit about yourself?
    \item Why did you start using LLMs? How long have you used it? How did you gain knowledge in prompting LLMs?  
    \item Have you used LLMs to assist your work as an instructor (e.g., syllabus creation, assignment/essay grading, facilitating in-class discussions), or a student (e.g., assignment, essay, Q\&A), or a professional (e.g., writing code)? How? Could you give me a concrete example? 
    \item What role do you and LLMs play, respectively, in the collaboration? How do you address LLMs?
    \item What is your overall experience and perception of LLM use? 
\end{itemize}

\noindent \textbf{Ethical Concerns regarding LLM Use}
\begin{itemize} 
    \item If you've used LLMs in educational/professional contexts, how has it helped you? (If not, why not?) Are there any risks you are particularly worried about regarding LLM use in educational/professional contexts?
    \item (Inaccurate answer) Are LLMs' responses always accurate for you? How do you think factual errors will affect students or professors/users?
    \item (Hallucination) ``Hallucination'' refers to a phenomenon where LLMs generate information or responses that are not accurate or grounded in reality. For example, it may generate a reference relevant to your prompt that doesn't actually exist. Have you encountered hallucinations? If so, could you give me an example? What do you think about its ethical implications in educational/professional contexts?
    \item (Bias) How do you see potential bias introduced by LLMs? (ChatGPT is known to write code to decide programmers are male and white) Have you experienced any bias? How do you think such bias will affect students and professors/users?
    \item (Privacy) Have you previously worried about the risk of privacy leakage when interacting with LLMs? Have you experienced any privacy leakage? How do you think it will affect students and professors/users? 
    \item (Academic integrity) Do you think LLM use can lead to academic integrity concerns (e.g., cheating in exams and plagiarism in research)? How to regulate academic integrity issues in the LLM era?
    % \item (Human autonomy) Are you afraid that LLMs will replace you as a professor/software engineer/... someday? Why? What will you do to prevent yourself from being replaced by it?
    \item Do you have other ethical concerns about LLM use?
    \item Do you use other AI and expert systems in general? How? What do you think is the difference between them and LLMs?
    \item What kind of education policy would you like to see regarding LLM use in higher education? Who should be responsible? Should teachers be educated? How should the line be drawn between useful AI and too-powerful or misused AI?
    \begin{itemize}
        \item (For professionals) What kind of regulation would you like to see regarding LLM use in your industry? How should the line be drawn between useful AI and too-powerful or misused AI? How should such regulation be different from education policy regarding LLM use? 
    \end{itemize}    
\end{itemize}

%% file: 9-demographic.tex
\newpage
\section{Demographic Information of The Participants}

\begin{table}[h]
  \caption{Demographic information of the participants. U indicates undergraduate students; M indicates master students; D indicates doctoral students; P indicates professors; I indicates industrial practitioners.}
  \label{demographic}
  \begin{tabular}{cccccl}
    \toprule
    Number & Major & Profession & Location & Gender\\
    \midrule
    U1 & Statistics & Undergraduate Student & US & M\\
    U2 & CS/Data Science & Undergraduate Student & US & F\\
    U3 & CS/Cognitive Science & Undergraduate Student & US & F\\
    U4 & IS & Undergraduate Student & US & M\\
    M1 & CS & Master Student & US & M\\
    M2 & IS & Master Student & US & F\\
    M3 & CS & Master Student & US & F\\
    M4 & IS & Master Student & US & M\\
    D1 & CS & Doctoral Student & US & M\\
    D2 & English & Doctoral Student & US & F\\
    D3 & CS & Doctoral Student & US & F\\
    D4 & IS & Doctoral Student & US & M\\
    P1 & IS & Professor & US & F\\
    P2 & IS & Professor & US & M\\
    P3 & IS & Professor & US & F\\
    P4 & IS & Professor & US & M\\
    I1 & - & Machine Learning Engineer & US & M\\
    I2 & - & Data Scientist & US & F\\
    I3 & - & AI Research Lead & US & F\\
    I4 & - & Machine Learning Lead & US & M\\
  \bottomrule
\end{tabular}
\end{table}

%% file: 0-facct.bbl
%%% -*-BibTeX-*-
%%% Do NOT edit. File created by BibTeX with style
%%% ACM-Reference-Format-Journals [18-Jan-2012].

\begin{thebibliography}{100}

%%% ====================================================================
%%% NOTE TO THE USER: you can override these defaults by providing
%%% customized versions of any of these macros before the \bibliography
%%% command.  Each of them MUST provide its own final punctuation,
%%% except for \shownote{}, \showDOI{}, and \showURL{}.  The latter two
%%% do not use final punctuation, in order to avoid confusing it with
%%% the Web address.
%%%
%%% To suppress output of a particular field, define its macro to expand
%%% to an empty string, or better, \unskip, like this:
%%%
%%% \newcommand{\showDOI}[1]{\unskip}   % LaTeX syntax
%%%
%%% \def \showDOI #1{\unskip}           % plain TeX syntax
%%%
%%% ====================================================================

\ifx \showCODEN    \undefined \def \showCODEN     #1{\unskip}     \fi
\ifx \showDOI      \undefined \def \showDOI       #1{#1}\fi
\ifx \showISBNx    \undefined \def \showISBNx     #1{\unskip}     \fi
\ifx \showISBNxiii \undefined \def \showISBNxiii  #1{\unskip}     \fi
\ifx \showISSN     \undefined \def \showISSN      #1{\unskip}     \fi
\ifx \showLCCN     \undefined \def \showLCCN      #1{\unskip}     \fi
\ifx \shownote     \undefined \def \shownote      #1{#1}          \fi
\ifx \showarticletitle \undefined \def \showarticletitle #1{#1}   \fi
\ifx \showURL      \undefined \def \showURL       {\relax}        \fi
% The following commands are used for tagged output and should be
% invisible to TeX
\providecommand\bibfield[2]{#2}
\providecommand\bibinfo[2]{#2}
\providecommand\natexlab[1]{#1}
\providecommand\showeprint[2][]{arXiv:#2}

\bibitem[Abdelghani et~al\mbox{.}(2023)]%
        {abdelghani_gpt-3-driven_2023}
\bibfield{author}{\bibinfo{person}{Rania Abdelghani}, \bibinfo{person}{Yen-Hsiang Wang}, \bibinfo{person}{Xingdi Yuan}, \bibinfo{person}{Tong Wang}, \bibinfo{person}{Pauline Lucas}, \bibinfo{person}{Hélène Sauzéon}, {and} \bibinfo{person}{Pierre-Yves Oudeyer}.} \bibinfo{year}{2023}\natexlab{}.
\newblock \showarticletitle{GPT-3-driven pedagogical agents for training children’s curious question-asking skills}.
\newblock \bibinfo{journal}{\emph{International Journal of Artificial Intelligence in Education}} (\bibinfo{date}{June} \bibinfo{year}{2023}).
\newblock
\showISSN{1560-4292, 1560-4306}
\urldef\tempurl%
\url{https://doi.org/10.1007/s40593-023-00340-7}
\showDOI{\tempurl}
\newblock
\shownote{arXiv:2211.14228 [cs]}.


\bibitem[Ahmad et~al\mbox{.}(2023)]%
        {ahmad2023towards}
\bibfield{author}{\bibinfo{person}{Aakash Ahmad}, \bibinfo{person}{Muhammad Waseem}, \bibinfo{person}{Peng Liang}, \bibinfo{person}{Mahdi Fahmideh}, \bibinfo{person}{Mst~Shamima Aktar}, {and} \bibinfo{person}{Tommi Mikkonen}.} \bibinfo{year}{2023}\natexlab{}.
\newblock \showarticletitle{Towards human-bot collaborative software architecting with chatgpt}. In \bibinfo{booktitle}{\emph{Proceedings of the 27th International Conference on Evaluation and Assessment in Software Engineering}}. \bibinfo{pages}{279--285}.
\newblock


\bibitem[Aithal and Aithal(2023)]%
        {aithal2023application}
\bibfield{author}{\bibinfo{person}{PS Aithal} {and} \bibinfo{person}{Shubhrajyosna Aithal}.} \bibinfo{year}{2023}\natexlab{}.
\newblock \showarticletitle{Application of ChatGPT in Higher Education and Research--A Futuristic Analysis}.
\newblock \bibinfo{journal}{\emph{International Journal of Applied Engineering and Management Letters (IJAEML)}} \bibinfo{volume}{7}, \bibinfo{number}{3} (\bibinfo{year}{2023}), \bibinfo{pages}{168--194}.
\newblock


\bibitem[Alkaissi and McFarlane(2023)]%
        {alkaissi2023artificial}
\bibfield{author}{\bibinfo{person}{Hussam Alkaissi} {and} \bibinfo{person}{Samy~I McFarlane}.} \bibinfo{year}{2023}\natexlab{}.
\newblock \showarticletitle{Artificial hallucinations in ChatGPT: implications in scientific writing}.
\newblock \bibinfo{journal}{\emph{Cureus}} \bibinfo{volume}{15}, \bibinfo{number}{2} (\bibinfo{year}{2023}).
\newblock


\bibitem[Ansari et~al\mbox{.}(2023)]%
        {ansari2023mapping}
\bibfield{author}{\bibinfo{person}{Aisha~Naz Ansari}, \bibinfo{person}{Sohail Ahmad}, {and} \bibinfo{person}{Sadia~Muzaffar Bhutta}.} \bibinfo{year}{2023}\natexlab{}.
\newblock \showarticletitle{Mapping the global evidence around the use of ChatGPT in higher education: A systematic scoping review}.
\newblock \bibinfo{journal}{\emph{Education and Information Technologies}} (\bibinfo{year}{2023}), \bibinfo{pages}{1--41}.
\newblock


\bibitem[Arias~Sosa and Godow(2023)]%
        {arias2023comparing}
\bibfield{author}{\bibinfo{person}{Elissa Arias~Sosa} {and} \bibinfo{person}{Marco Godow}.} \bibinfo{year}{2023}\natexlab{}.
\newblock \bibinfo{title}{Comparing Google and ChatGPT as Assistive Tools for Students in Solving Programming Exercises}.
\newblock
\newblock


\bibitem[Azoulay et~al\mbox{.}(2023)]%
        {azoulay2023let}
\bibfield{author}{\bibinfo{person}{Rina Azoulay}, \bibinfo{person}{Tirzta Hirst}, {and} \bibinfo{person}{Shulamit Reches}.} \bibinfo{year}{2023}\natexlab{}.
\newblock \showarticletitle{Let’s Do It Ourselves: Ensuring Academic Integrity in the Age of ChatGPT and Beyond}.
\newblock \bibinfo{journal}{\emph{Authorea Preprints}} (\bibinfo{year}{2023}).
\newblock


\bibitem[Backstr{\"o}m and Kihlert(2023)]%
        {backstrom2023code}
\bibfield{author}{\bibinfo{person}{Oscar Backstr{\"o}m} {and} \bibinfo{person}{Annie Kihlert}.} \bibinfo{year}{2023}\natexlab{}.
\newblock \bibinfo{title}{Code Quality and Large Language Models in Computer Science Education: Enhancing student-written code through ChatGPT}.
\newblock
\newblock


\bibitem[Bao({[n.\,d.]})]%
        {bao_can_nodate}
\bibfield{author}{\bibinfo{person}{Minhui Bao}.} \bibinfo{year}{[n.\,d.]}\natexlab{}.
\newblock \showarticletitle{Can Home Use of Speech-Enabled Artificial Intelligence Mitigate Foreign Language Anxiety – Investigation of a Concept}.
\newblock \bibinfo{journal}{\emph{Arab World English Journal (AWEJ)}}  \bibinfo{volume}{5} (\bibinfo{year}{[n.\,d.]}), \bibinfo{pages}{28–40}.
\newblock


\bibitem[Barbas et~al\mbox{.}(2023)]%
        {barbas2023importance}
\bibfield{author}{\bibinfo{person}{Maria~Potes Barbas}, \bibinfo{person}{Andreia~Teles Vieira}, {and} \bibinfo{person}{Paulo~Duarte Branco}.} \bibinfo{year}{2023}\natexlab{}.
\newblock \showarticletitle{The Importance of Chat GPT Training for Higher Education: Case Study}. In \bibinfo{booktitle}{\emph{International Conference on Design and Digital Communication}}. Springer, \bibinfo{pages}{695--705}.
\newblock


\bibitem[Baskara et~al\mbox{.}(2023)]%
        {baskara2023chatgpt}
\bibfield{author}{\bibinfo{person}{FX~Risang Baskara}, \bibinfo{person}{Anindita~Dewangga Puri}, {and} \bibinfo{person}{Annisa~Radista Wardhani}.} \bibinfo{year}{2023}\natexlab{}.
\newblock \showarticletitle{ChatGPT and the Pedagogical Challenge: Unveiling the Impact on Early-Career Academics in Higher Education}.
\newblock \bibinfo{journal}{\emph{Indonesian Journal on Learning and Advanced Education (IJOLAE)}} \bibinfo{volume}{5}, \bibinfo{number}{3} (\bibinfo{year}{2023}), \bibinfo{pages}{311--322}.
\newblock


\bibitem[Becker et~al\mbox{.}(2023)]%
        {becker_programming_2023}
\bibfield{author}{\bibinfo{person}{Brett~A. Becker}, \bibinfo{person}{Paul Denny}, \bibinfo{person}{James Finnie-Ansley}, \bibinfo{person}{Andrew Luxton-Reilly}, \bibinfo{person}{James Prather}, {and} \bibinfo{person}{Eddie~Antonio Santos}.} \bibinfo{year}{2023}\natexlab{}.
\newblock \showarticletitle{Programming {Is} {Hard} - {Or} at {Least} {It} {Used} to {Be}: {Educational} {Opportunities} and {Challenges} of {AI} {Code} {Generation}}. In \bibinfo{booktitle}{\emph{Proceedings of the 54th {ACM} {Technical} {Symposium} on {Computer} {Science} {Education} {V}. 1}}. \bibinfo{publisher}{ACM}, \bibinfo{address}{Toronto ON Canada}, \bibinfo{pages}{500--506}.
\newblock
\showISBNx{978-1-4503-9431-4}
\urldef\tempurl%
\url{https://doi.org/10.1145/3545945.3569759}
\showDOI{\tempurl}


\bibitem[Bender et~al\mbox{.}(2021)]%
        {bender_dangers_2021}
\bibfield{author}{\bibinfo{person}{Emily~M. Bender}, \bibinfo{person}{Timnit Gebru}, \bibinfo{person}{Angelina McMillan-Major}, {and} \bibinfo{person}{Shmargaret Shmitchell}.} \bibinfo{year}{2021}\natexlab{}.
\newblock \showarticletitle{On the {Dangers} of {Stochastic} {Parrots}: {Can} {Language} {Models} {Be} {Too} {Big}?}. In \bibinfo{booktitle}{\emph{Proceedings of the 2021 {ACM} {Conference} on {Fairness}, {Accountability}, and {Transparency}}}. \bibinfo{publisher}{ACM}, \bibinfo{address}{Virtual Event Canada}, \bibinfo{pages}{610--623}.
\newblock
\showISBNx{978-1-4503-8309-7}
\urldef\tempurl%
\url{https://doi.org/10.1145/3442188.3445922}
\showDOI{\tempurl}


\bibitem[Bhat et~al\mbox{.}(2022)]%
        {bhat_towards_2022}
\bibfield{author}{\bibinfo{person}{Shravya Bhat}, \bibinfo{person}{Huy Nguyen}, \bibinfo{person}{Steven Moore}, \bibinfo{person}{John Stamper}, \bibinfo{person}{Majd Sakr}, {and} \bibinfo{person}{Eric Nyberg}.} \bibinfo{year}{2022}\natexlab{}.
\newblock \showarticletitle{Towards Automated Generation and Evaluation of Questions in Educational Domains}. In \bibinfo{booktitle}{\emph{Proceedings of the 15th International Conference on Educational Data Mining}}. \bibinfo{publisher}{International Educational Data Mining Society}, \bibinfo{address}{Durham, United Kingdom}, \bibinfo{pages}{701–704}.
\newblock
\showISBNx{9781733673631}
\urldef\tempurl%
\url{https://doi.org/10.5281/zenodo.6853085}
\showDOI{\tempurl}


\bibitem[Bird et~al\mbox{.}(2022)]%
        {bird_taking_2022}
\bibfield{author}{\bibinfo{person}{Christian Bird}, \bibinfo{person}{Denae Ford}, \bibinfo{person}{Thomas Zimmermann}, \bibinfo{person}{Nicole Forsgren}, \bibinfo{person}{Eirini Kalliamvakou}, \bibinfo{person}{Travis Lowdermilk}, {and} \bibinfo{person}{Idan Gazit}.} \bibinfo{year}{2022}\natexlab{}.
\newblock \showarticletitle{Taking {Flight} with {Copilot}: {Early} insights and opportunities of {AI}-powered pair-programming tools}.
\newblock \bibinfo{journal}{\emph{Queue}} \bibinfo{volume}{20}, \bibinfo{number}{6} (\bibinfo{date}{Dec.} \bibinfo{year}{2022}), \bibinfo{pages}{35--57}.
\newblock
\showISSN{1542-7730, 1542-7749}
\urldef\tempurl%
\url{https://doi.org/10.1145/3582083}
\showDOI{\tempurl}


\bibitem[Biswas(2023)]%
        {biswas2023role}
\bibfield{author}{\bibinfo{person}{Som Biswas}.} \bibinfo{year}{2023}\natexlab{}.
\newblock \showarticletitle{Role of ChatGPT in Computer Programming.: ChatGPT in Computer Programming.}
\newblock \bibinfo{journal}{\emph{Mesopotamian Journal of Computer Science}}  \bibinfo{volume}{2023} (\bibinfo{year}{2023}), \bibinfo{pages}{8--16}.
\newblock


\bibitem[Bordt and von Luxburg(2023)]%
        {bordt2023chatgpt}
\bibfield{author}{\bibinfo{person}{Sebastian Bordt} {and} \bibinfo{person}{Ulrike von Luxburg}.} \bibinfo{year}{2023}\natexlab{}.
\newblock \showarticletitle{Chatgpt participates in a computer science exam}.
\newblock \bibinfo{journal}{\emph{arXiv preprint arXiv:2303.09461}} (\bibinfo{year}{2023}).
\newblock


\bibitem[Braun and Clarke(2012)]%
        {braun2012thematic}
\bibfield{author}{\bibinfo{person}{Virginia Braun} {and} \bibinfo{person}{Victoria Clarke}.} \bibinfo{year}{2012}\natexlab{}.
\newblock \bibinfo{booktitle}{\emph{Thematic analysis.}}
\newblock \bibinfo{publisher}{American Psychological Association}.
\newblock


\bibitem[Chaudhry et~al\mbox{.}(2023)]%
        {chaudhry2023time}
\bibfield{author}{\bibinfo{person}{Iffat~Sabir Chaudhry}, \bibinfo{person}{Sayed Ahmad~M Sarwary}, \bibinfo{person}{Ghaleb~A El~Refae}, {and} \bibinfo{person}{Habib Chabchoub}.} \bibinfo{year}{2023}\natexlab{}.
\newblock \showarticletitle{Time to Revisit Existing Student’s Performance Evaluation Approach in Higher Education Sector in a New Era of ChatGPT—A Case Study}.
\newblock \bibinfo{journal}{\emph{Cogent Education}} \bibinfo{volume}{10}, \bibinfo{number}{1} (\bibinfo{year}{2023}), \bibinfo{pages}{2210461}.
\newblock


\bibitem[Chen et~al\mbox{.}(2021)]%
        {chen_evaluating_2021}
\bibfield{author}{\bibinfo{person}{Mark Chen}, \bibinfo{person}{Jerry Tworek}, \bibinfo{person}{Heewoo Jun}, \bibinfo{person}{Qiming Yuan}, \bibinfo{person}{Henrique Ponde de~Oliveira Pinto}, \bibinfo{person}{Jared Kaplan}, \bibinfo{person}{Harri Edwards}, \bibinfo{person}{Yuri Burda}, \bibinfo{person}{Nicholas Joseph}, \bibinfo{person}{Greg Brockman}, \bibinfo{person}{Alex Ray}, \bibinfo{person}{Raul Puri}, \bibinfo{person}{Gretchen Krueger}, \bibinfo{person}{Michael Petrov}, \bibinfo{person}{Heidy Khlaaf}, \bibinfo{person}{Girish Sastry}, \bibinfo{person}{Pamela Mishkin}, \bibinfo{person}{Brooke Chan}, \bibinfo{person}{Scott Gray}, \bibinfo{person}{Nick Ryder}, \bibinfo{person}{Mikhail Pavlov}, \bibinfo{person}{Alethea Power}, \bibinfo{person}{Lukasz Kaiser}, \bibinfo{person}{Mohammad Bavarian}, \bibinfo{person}{Clemens Winter}, \bibinfo{person}{Philippe Tillet}, \bibinfo{person}{Felipe~Petroski Such}, \bibinfo{person}{Dave Cummings}, \bibinfo{person}{Matthias Plappert}, \bibinfo{person}{Fotios Chantzis},
  \bibinfo{person}{Elizabeth Barnes}, \bibinfo{person}{Ariel Herbert-Voss}, \bibinfo{person}{William~Hebgen Guss}, \bibinfo{person}{Alex Nichol}, \bibinfo{person}{Alex Paino}, \bibinfo{person}{Nikolas Tezak}, \bibinfo{person}{Jie Tang}, \bibinfo{person}{Igor Babuschkin}, \bibinfo{person}{Suchir Balaji}, \bibinfo{person}{Shantanu Jain}, \bibinfo{person}{William Saunders}, \bibinfo{person}{Christopher Hesse}, \bibinfo{person}{Andrew~N. Carr}, \bibinfo{person}{Jan Leike}, \bibinfo{person}{Josh Achiam}, \bibinfo{person}{Vedant Misra}, \bibinfo{person}{Evan Morikawa}, \bibinfo{person}{Alec Radford}, \bibinfo{person}{Matthew Knight}, \bibinfo{person}{Miles Brundage}, \bibinfo{person}{Mira Murati}, \bibinfo{person}{Katie Mayer}, \bibinfo{person}{Peter Welinder}, \bibinfo{person}{Bob McGrew}, \bibinfo{person}{Dario Amodei}, \bibinfo{person}{Sam McCandlish}, \bibinfo{person}{Ilya Sutskever}, {and} \bibinfo{person}{Wojciech Zaremba}.} \bibinfo{year}{2021}\natexlab{}.
\newblock \bibinfo{title}{Evaluating {Large} {Language} {Models} {Trained} on {Code}}.
\newblock
\newblock
\urldef\tempurl%
\url{http://arxiv.org/abs/2107.03374}
\showURL{%
\tempurl}
\newblock
\shownote{arXiv:2107.03374 [cs]}.


\bibitem[Cohney et~al\mbox{.}(2021)]%
        {cohney2021virtual}
\bibfield{author}{\bibinfo{person}{Shaanan Cohney}, \bibinfo{person}{Ross Teixeira}, \bibinfo{person}{Anne Kohlbrenner}, \bibinfo{person}{Arvind Narayanan}, \bibinfo{person}{Mihir Kshirsagar}, \bibinfo{person}{Yan Shvartzshnaider}, {and} \bibinfo{person}{Madelyn Sanfilippo}.} \bibinfo{year}{2021}\natexlab{}.
\newblock \showarticletitle{Virtual Classrooms and Real Harms: Remote Learning at $\{$US$\}$. Universities}. In \bibinfo{booktitle}{\emph{Seventeenth Symposium on Usable Privacy and Security (SOUPS 2021)}}. \bibinfo{pages}{653--674}.
\newblock


\bibitem[Cotton et~al\mbox{.}(2023)]%
        {cotton2023chatting}
\bibfield{author}{\bibinfo{person}{Debby~RE Cotton}, \bibinfo{person}{Peter~A Cotton}, {and} \bibinfo{person}{J~Reuben Shipway}.} \bibinfo{year}{2023}\natexlab{}.
\newblock \showarticletitle{Chatting and cheating: Ensuring academic integrity in the era of ChatGPT}.
\newblock \bibinfo{journal}{\emph{Innovations in Education and Teaching International}} (\bibinfo{year}{2023}), \bibinfo{pages}{1--12}.
\newblock


\bibitem[Currie(2023)]%
        {currie2023academic}
\bibfield{author}{\bibinfo{person}{Geoffrey~M Currie}.} \bibinfo{year}{2023}\natexlab{}.
\newblock \showarticletitle{Academic integrity and artificial intelligence: is ChatGPT hype, hero or heresy?}. In \bibinfo{booktitle}{\emph{Seminars in Nuclear Medicine}}. Elsevier.
\newblock


\bibitem[Dai et~al\mbox{.}(2023)]%
        {dai2023reconceptualizing}
\bibfield{author}{\bibinfo{person}{Yun Dai}, \bibinfo{person}{Ang Liu}, {and} \bibinfo{person}{Cher~Ping Lim}.} \bibinfo{year}{2023}\natexlab{}.
\newblock \showarticletitle{Reconceptualizing ChatGPT and generative AI as a student-driven innovation in higher education}.
\newblock  (\bibinfo{year}{2023}).
\newblock


\bibitem[Dansdill et~al\mbox{.}(2008)]%
        {dansdill2008exposing}
\bibfield{author}{\bibinfo{person}{Timothy Dansdill}, \bibinfo{person}{Mark~E Hoffman}, {and} \bibinfo{person}{David~S Herscovici}.} \bibinfo{year}{2008}\natexlab{}.
\newblock \showarticletitle{Exposing gaps, exploring legacies: paradoxes of writing use in computing education}.
\newblock \bibinfo{journal}{\emph{Journal of Computing Sciences in Colleges}} \bibinfo{volume}{23}, \bibinfo{number}{5} (\bibinfo{year}{2008}), \bibinfo{pages}{24--33}.
\newblock


\bibitem[Dempere et~al\mbox{.}(2023)]%
        {dempere2023impact}
\bibfield{author}{\bibinfo{person}{Juan Dempere}, \bibinfo{person}{Kennedy~Prince Modugu}, \bibinfo{person}{Allam Hesham}, {and} \bibinfo{person}{Lakshmana Ramasamy}.} \bibinfo{year}{2023}\natexlab{}.
\newblock \showarticletitle{The impact of ChatGPT on higher education}.
\newblock \bibinfo{journal}{\emph{Dempere J, Modugu K, Hesham A and Ramasamy LK (2023) The impact of ChatGPT on higher education. Front. Educ}}  \bibinfo{volume}{8} (\bibinfo{year}{2023}), \bibinfo{pages}{1206936}.
\newblock


\bibitem[Di~Fede et~al\mbox{.}(2022)]%
        {di2022idea}
\bibfield{author}{\bibinfo{person}{Giulia Di~Fede}, \bibinfo{person}{Davide Rocchesso}, \bibinfo{person}{Steven~P Dow}, {and} \bibinfo{person}{Salvatore Andolina}.} \bibinfo{year}{2022}\natexlab{}.
\newblock \showarticletitle{The Idea Machine: LLM-based Expansion, Rewriting, Combination, and Suggestion of Ideas}. In \bibinfo{booktitle}{\emph{Proceedings of the 14th Conference on Creativity and Cognition}}. \bibinfo{pages}{623--627}.
\newblock


\bibitem[Eager and Brunton(2023)]%
        {eager2023prompting}
\bibfield{author}{\bibinfo{person}{Bronwyn Eager} {and} \bibinfo{person}{Ryan Brunton}.} \bibinfo{year}{2023}\natexlab{}.
\newblock \showarticletitle{Prompting higher education towards AI-augmented teaching and learning practice}.
\newblock \bibinfo{journal}{\emph{Journal of University Teaching \& Learning Practice}} \bibinfo{volume}{20}, \bibinfo{number}{5} (\bibinfo{year}{2023}), \bibinfo{pages}{02}.
\newblock


\bibitem[Felkner et~al\mbox{.}(2023)]%
        {felkner2023winoqueer}
\bibfield{author}{\bibinfo{person}{Virginia~K Felkner}, \bibinfo{person}{Ho-Chun~Herbert Chang}, \bibinfo{person}{Eugene Jang}, {and} \bibinfo{person}{Jonathan May}.} \bibinfo{year}{2023}\natexlab{}.
\newblock \showarticletitle{Winoqueer: A community-in-the-loop benchmark for anti-lgbtq+ bias in large language models}.
\newblock \bibinfo{journal}{\emph{arXiv preprint arXiv:2306.15087}} (\bibinfo{year}{2023}).
\newblock


\bibitem[Firaina and Sulisworo(2023)]%
        {firaina2023exploring}
\bibfield{author}{\bibinfo{person}{Radha Firaina} {and} \bibinfo{person}{Dwi Sulisworo}.} \bibinfo{year}{2023}\natexlab{}.
\newblock \showarticletitle{Exploring the usage of ChatGPT in higher education: Frequency and impact on productivity}.
\newblock \bibinfo{journal}{\emph{Buletin Edukasi Indonesia}} \bibinfo{volume}{2}, \bibinfo{number}{01} (\bibinfo{year}{2023}), \bibinfo{pages}{39--46}.
\newblock


\bibitem[Fowler et~al\mbox{.}(2023)]%
        {fowler2023first}
\bibfield{author}{\bibinfo{person}{Samuel Fowler}, \bibinfo{person}{Malgorzata Korolkiewicz}, {and} \bibinfo{person}{Rebecca Marrone}.} \bibinfo{year}{2023}\natexlab{}.
\newblock \showarticletitle{First 100 days of ChatGPT at Australian universities: An analysis of policy landscape and media discussions about the role of AI in higher education}.
\newblock \bibinfo{journal}{\emph{Learning Letters}} (\bibinfo{year}{2023}).
\newblock


\bibitem[Fuchs(2023)]%
        {fuchs2023exploring}
\bibfield{author}{\bibinfo{person}{Kevin Fuchs}.} \bibinfo{year}{2023}\natexlab{}.
\newblock \showarticletitle{Exploring the opportunities and challenges of NLP models in higher education: is Chat GPT a blessing or a curse?}. In \bibinfo{booktitle}{\emph{Frontiers in Education}}, Vol.~\bibinfo{volume}{8}. Frontiers, \bibinfo{pages}{1166682}.
\newblock


\bibitem[Fui-Hoon~Nah et~al\mbox{.}(2023)]%
        {fui2023generative}
\bibfield{author}{\bibinfo{person}{Fiona Fui-Hoon~Nah}, \bibinfo{person}{Ruilin Zheng}, \bibinfo{person}{Jingyuan Cai}, \bibinfo{person}{Keng Siau}, {and} \bibinfo{person}{Langtao Chen}.} \bibinfo{year}{2023}\natexlab{}.
\newblock \bibinfo{title}{Generative AI and ChatGPT: Applications, challenges, and AI-human collaboration}.
\newblock , \bibinfo{numpages}{277--304}~pages.
\newblock


\bibitem[George et~al\mbox{.}(2023)]%
        {george2023chatgpt}
\bibfield{author}{\bibinfo{person}{A~Shaji George}, \bibinfo{person}{AS~Hovan George}, {and} \bibinfo{person}{AS~Gabrio Martin}.} \bibinfo{year}{2023}\natexlab{}.
\newblock \showarticletitle{ChatGPT and the Future of Work: A Comprehensive Analysis of AI's Impact on Jobs and Employment}.
\newblock \bibinfo{journal}{\emph{Partners Universal International Innovation Journal}} \bibinfo{volume}{1}, \bibinfo{number}{3} (\bibinfo{year}{2023}), \bibinfo{pages}{154--186}.
\newblock


\bibitem[Gero et~al\mbox{.}(2022)]%
        {gero_sparks_2022}
\bibfield{author}{\bibinfo{person}{Katy~Ilonka Gero}, \bibinfo{person}{Vivian Liu}, {and} \bibinfo{person}{Lydia Chilton}.} \bibinfo{year}{2022}\natexlab{}.
\newblock \showarticletitle{Sparks: Inspiration for Science Writing using Language Models}. In \bibinfo{booktitle}{\emph{Proceedings of the 2022 ACM Designing Interactive Systems Conference}} \emph{(\bibinfo{series}{DIS ’22})}. \bibinfo{publisher}{Association for Computing Machinery}, \bibinfo{address}{New York, NY, USA}, \bibinfo{pages}{1002–1019}.
\newblock
\showISBNx{978-1-4503-9358-4}
\urldef\tempurl%
\url{https://doi.org/10.1145/3532106.3533533}
\showDOI{\tempurl}


\bibitem[Gill et~al\mbox{.}(2024)]%
        {gill2024transformative}
\bibfield{author}{\bibinfo{person}{Sukhpal~Singh Gill}, \bibinfo{person}{Minxian Xu}, \bibinfo{person}{Panos Patros}, \bibinfo{person}{Huaming Wu}, \bibinfo{person}{Rupinder Kaur}, \bibinfo{person}{Kamalpreet Kaur}, \bibinfo{person}{Stephanie Fuller}, \bibinfo{person}{Manmeet Singh}, \bibinfo{person}{Priyansh Arora}, \bibinfo{person}{Ajith~Kumar Parlikad}, {et~al\mbox{.}}} \bibinfo{year}{2024}\natexlab{}.
\newblock \showarticletitle{Transformative effects of ChatGPT on modern education: Emerging Era of AI Chatbots}.
\newblock \bibinfo{journal}{\emph{Internet of Things and Cyber-Physical Systems}}  \bibinfo{volume}{4} (\bibinfo{year}{2024}), \bibinfo{pages}{19--23}.
\newblock


\bibitem[Gupta et~al\mbox{.}(2023)]%
        {gupta2023chatgpt}
\bibfield{author}{\bibinfo{person}{Maanak Gupta}, \bibinfo{person}{CharanKumar Akiri}, \bibinfo{person}{Kshitiz Aryal}, \bibinfo{person}{Eli Parker}, {and} \bibinfo{person}{Lopamudra Praharaj}.} \bibinfo{year}{2023}\natexlab{}.
\newblock \showarticletitle{From chatgpt to threatgpt: Impact of generative ai in cybersecurity and privacy}.
\newblock \bibinfo{journal}{\emph{IEEE Access}} (\bibinfo{year}{2023}).
\newblock


\bibitem[Halaweh(2023)]%
        {halaweh2023chatgpt}
\bibfield{author}{\bibinfo{person}{Mohanad Halaweh}.} \bibinfo{year}{2023}\natexlab{}.
\newblock \showarticletitle{ChatGPT in education: Strategies for responsible implementation}.
\newblock  (\bibinfo{year}{2023}).
\newblock


\bibitem[Hasanein and Sobaih(2023)]%
        {hasanein2023drivers}
\bibfield{author}{\bibinfo{person}{Ahmed~M Hasanein} {and} \bibinfo{person}{Abu Elnasr~E Sobaih}.} \bibinfo{year}{2023}\natexlab{}.
\newblock \showarticletitle{Drivers and Consequences of ChatGPT Use in Higher Education: Key Stakeholder Perspectives}.
\newblock \bibinfo{journal}{\emph{European Journal of Investigation in Health, Psychology and Education}} \bibinfo{volume}{13}, \bibinfo{number}{11} (\bibinfo{year}{2023}), \bibinfo{pages}{2599--2614}.
\newblock


\bibitem[H{\"o}rnemalm(2023)]%
        {hornemalm2023chatgpt}
\bibfield{author}{\bibinfo{person}{Adam H{\"o}rnemalm}.} \bibinfo{year}{2023}\natexlab{}.
\newblock \bibinfo{title}{ChatGPT as a Software Development Tool: The Future of Development}.
\newblock
\newblock


\bibitem[Hu and Twidale(2023)]%
        {hu2023scoping}
\bibfield{author}{\bibinfo{person}{Xinhui Hu} {and} \bibinfo{person}{Michael Twidale}.} \bibinfo{year}{2023}\natexlab{}.
\newblock \showarticletitle{A Scoping Review of Mental Model Research in HCI from 2010 to 2021}. In \bibinfo{booktitle}{\emph{International Conference on Human-Computer Interaction}}. Springer, \bibinfo{pages}{101--125}.
\newblock


\bibitem[Huallpa et~al\mbox{.}(2023)]%
        {huallpa2023exploring}
\bibfield{author}{\bibinfo{person}{Jorge~Jinchu{\~n}a Huallpa} {et~al\mbox{.}}} \bibinfo{year}{2023}\natexlab{}.
\newblock \showarticletitle{Exploring the ethical considerations of using Chat GPT in university education}.
\newblock \bibinfo{journal}{\emph{Periodicals of Engineering and Natural Sciences}} \bibinfo{volume}{11}, \bibinfo{number}{4} (\bibinfo{year}{2023}), \bibinfo{pages}{105--115}.
\newblock


\bibitem[Huang et~al\mbox{.}(2023)]%
        {huang2023bias}
\bibfield{author}{\bibinfo{person}{Dong Huang}, \bibinfo{person}{Qingwen Bu}, \bibinfo{person}{Jie Zhang}, \bibinfo{person}{Xiaofei Xie}, \bibinfo{person}{Junjie Chen}, {and} \bibinfo{person}{Heming Cui}.} \bibinfo{year}{2023}\natexlab{}.
\newblock \showarticletitle{Bias assessment and mitigation in llm-based code generation}.
\newblock \bibinfo{journal}{\emph{arXiv preprint arXiv:2309.14345}} (\bibinfo{year}{2023}).
\newblock


\bibitem[Huang et~al\mbox{.}(2024)]%
        {huang_bias_2024}
\bibfield{author}{\bibinfo{person}{Dong Huang}, \bibinfo{person}{Qingwen Bu}, \bibinfo{person}{Jie Zhang}, \bibinfo{person}{Xiaofei Xie}, \bibinfo{person}{Junjie Chen}, {and} \bibinfo{person}{Heming Cui}.} \bibinfo{year}{2024}\natexlab{}.
\newblock \bibinfo{title}{Bias {Testing} and {Mitigation} in {LLM}-based {Code} {Generation}}.
\newblock
\newblock
\urldef\tempurl%
\url{http://arxiv.org/abs/2309.14345}
\showURL{%
\tempurl}
\newblock
\shownote{arXiv:2309.14345 [cs]}.


\bibitem[Jacobsen and Weber(2023)]%
        {jacobsen2023promises}
\bibfield{author}{\bibinfo{person}{Lucas~Jasper Jacobsen} {and} \bibinfo{person}{Kira~Elena Weber}.} \bibinfo{year}{2023}\natexlab{}.
\newblock \showarticletitle{The Promises and Pitfalls of ChatGPT as a Feedback Provider in Higher Education: An Exploratory Study of Prompt Engineering and the Quality of AI-Driven Feedback}.
\newblock  (\bibinfo{year}{2023}).
\newblock


\bibitem[Jalil et~al\mbox{.}(2023)]%
        {jalil2023chatgpt}
\bibfield{author}{\bibinfo{person}{Sajed Jalil}, \bibinfo{person}{Suzzana Rafi}, \bibinfo{person}{Thomas~D LaToza}, \bibinfo{person}{Kevin Moran}, {and} \bibinfo{person}{Wing Lam}.} \bibinfo{year}{2023}\natexlab{}.
\newblock \showarticletitle{Chatgpt and software testing education: Promises \& perils}. In \bibinfo{booktitle}{\emph{2023 IEEE International Conference on Software Testing, Verification and Validation Workshops (ICSTW)}}. IEEE, \bibinfo{pages}{4130--4137}.
\newblock


\bibitem[Javaid et~al\mbox{.}(2023)]%
        {javaid2023unlocking}
\bibfield{author}{\bibinfo{person}{Mohd Javaid}, \bibinfo{person}{Abid Haleem}, \bibinfo{person}{Ravi~Pratap Singh}, \bibinfo{person}{Shahbaz Khan}, {and} \bibinfo{person}{Ibrahim~Haleem Khan}.} \bibinfo{year}{2023}\natexlab{}.
\newblock \showarticletitle{Unlocking the opportunities through ChatGPT Tool towards ameliorating the education system}.
\newblock \bibinfo{journal}{\emph{BenchCouncil Transactions on Benchmarks, Standards and Evaluations}} \bibinfo{volume}{3}, \bibinfo{number}{2} (\bibinfo{year}{2023}), \bibinfo{pages}{100115}.
\newblock


\bibitem[Jukiewicz({[n.\,d.]})]%
        {jukiewiczfuture}
\bibfield{author}{\bibinfo{person}{Marcin Jukiewicz}.} \bibinfo{year}{[n.\,d.]}\natexlab{}.
\newblock \showarticletitle{The Future of Grading Programming Assignments in Education: The Role of ChatGPT in Automating the Assessment and Feedback Process}.
\newblock  (\bibinfo{year}{[n.\,d.]}).
\newblock


\bibitem[Kasneci et~al\mbox{.}(2023)]%
        {kasneci2023chatgpt}
\bibfield{author}{\bibinfo{person}{Enkelejda Kasneci}, \bibinfo{person}{Kathrin Se{\ss}ler}, \bibinfo{person}{Stefan K{\"u}chemann}, \bibinfo{person}{Maria Bannert}, \bibinfo{person}{Daryna Dementieva}, \bibinfo{person}{Frank Fischer}, \bibinfo{person}{Urs Gasser}, \bibinfo{person}{Georg Groh}, \bibinfo{person}{Stephan G{\"u}nnemann}, \bibinfo{person}{Eyke H{\"u}llermeier}, {et~al\mbox{.}}} \bibinfo{year}{2023}\natexlab{}.
\newblock \showarticletitle{ChatGPT for good? On opportunities and challenges of large language models for education}.
\newblock \bibinfo{journal}{\emph{Learning and individual differences}}  \bibinfo{volume}{103} (\bibinfo{year}{2023}), \bibinfo{pages}{102274}.
\newblock


\bibitem[Khaddage(2023)]%
        {khaddage2023towards}
\bibfield{author}{\bibinfo{person}{Ferial Khaddage}.} \bibinfo{year}{2023}\natexlab{}.
\newblock \showarticletitle{Towards an Innovative Strategy for ChatGPT in Higher Education “Respond, Reimagine, Recreate, \& Reform”}. In \bibinfo{booktitle}{\emph{EdMedia+ Innovate Learning}}. Association for the Advancement of Computing in Education (AACE), \bibinfo{pages}{274--279}.
\newblock


\bibitem[Kilhoffer et~al\mbox{.}(2023a)]%
        {kilhoffer2023ai}
\bibfield{author}{\bibinfo{person}{Zachary Kilhoffer}, \bibinfo{person}{Anita Nlkolich}, \bibinfo{person}{Madelyn~Rose Sanfilippo}, {and} \bibinfo{person}{Zhixuan Zhou}.} \bibinfo{year}{2023}\natexlab{a}.
\newblock \showarticletitle{AI Accountability Policy}.
\newblock \bibinfo{journal}{\emph{NTIA-2023-0005-0810}} (\bibinfo{year}{2023}).
\newblock


\bibitem[Kilhoffer et~al\mbox{.}(2023b)]%
        {kilhoffer2023technical}
\bibfield{author}{\bibinfo{person}{Zachary Kilhoffer}, \bibinfo{person}{Zhixuan Zhou}, \bibinfo{person}{Firmiana Wang}, \bibinfo{person}{Fahad Tamton}, \bibinfo{person}{Yun Huang}, \bibinfo{person}{Pilyoung Kim}, \bibinfo{person}{Tom Yeh}, {and} \bibinfo{person}{Yang Wang}.} \bibinfo{year}{2023}\natexlab{b}.
\newblock \showarticletitle{“How technical do you get? I’m an English teacher”: Teaching and Learning Cybersecurity and AI Ethics in High School}. In \bibinfo{booktitle}{\emph{2023 IEEE Symposium on Security and Privacy (SP)}}. IEEE, \bibinfo{pages}{2032--2032}.
\newblock


\bibitem[Kiryakova and Angelova(2023)]%
        {kiryakova2023chatgpt}
\bibfield{author}{\bibinfo{person}{Gabriela Kiryakova} {and} \bibinfo{person}{Nadezhda Angelova}.} \bibinfo{year}{2023}\natexlab{}.
\newblock \showarticletitle{ChatGPT—A Challenging Tool for the University Professors in Their Teaching Practice}.
\newblock \bibinfo{journal}{\emph{Education Sciences}} \bibinfo{volume}{13}, \bibinfo{number}{10} (\bibinfo{year}{2023}), \bibinfo{pages}{1056}.
\newblock


\bibitem[Kumar et~al\mbox{.}(2023)]%
        {kumar2023impact}
\bibfield{author}{\bibinfo{person}{Harsh Kumar}, \bibinfo{person}{Ilya Musabirov}, \bibinfo{person}{Mohi Reza}, \bibinfo{person}{Jiakai Shi}, \bibinfo{person}{Anastasia Kuzminykh}, \bibinfo{person}{Joseph~Jay Williams}, {and} \bibinfo{person}{Michael Liut}.} \bibinfo{year}{2023}\natexlab{}.
\newblock \showarticletitle{Impact of Guidance and Interaction Strategies for LLM Use on Learner Performance and Perception}.
\newblock \bibinfo{journal}{\emph{arXiv preprint arXiv:2310.13712}} (\bibinfo{year}{2023}).
\newblock


\bibitem[Laato et~al\mbox{.}(2023)]%
        {laato2023ai}
\bibfield{author}{\bibinfo{person}{Samuli Laato}, \bibinfo{person}{Benedikt Morschheuser}, \bibinfo{person}{Juho Hamari}, {and} \bibinfo{person}{Jari Bj{\"o}rne}.} \bibinfo{year}{2023}\natexlab{}.
\newblock \showarticletitle{AI-assisted learning with ChatGPT and large language models: Implications for higher education}. In \bibinfo{booktitle}{\emph{2023 IEEE International Conference on Advanced Learning Technologies (ICALT)}}. IEEE, \bibinfo{pages}{226--230}.
\newblock


\bibitem[Lakhani and Wolf(2003)]%
        {lakhani2003hackers}
\bibfield{author}{\bibinfo{person}{Karim~R Lakhani} {and} \bibinfo{person}{Robert~G Wolf}.} \bibinfo{year}{2003}\natexlab{}.
\newblock \showarticletitle{Why hackers do what they do: Understanding motivation and effort in free/open source software projects}.
\newblock \bibinfo{journal}{\emph{Open Source Software Projects (September 2003)}} (\bibinfo{year}{2003}).
\newblock


\bibitem[Lau and Guo(2023)]%
        {lau2023ban}
\bibfield{author}{\bibinfo{person}{Sam Lau} {and} \bibinfo{person}{Philip Guo}.} \bibinfo{year}{2023}\natexlab{}.
\newblock \showarticletitle{From" Ban it till we understand it" to" Resistance is futile": How university programming instructors plan to adapt as more students use AI code generation and explanation tools such as ChatGPT and GitHub Copilot}. In \bibinfo{booktitle}{\emph{Proceedings of the 2023 ACM Conference on International Computing Education Research-Volume 1}}. \bibinfo{pages}{106--121}.
\newblock


\bibitem[Lee and Soylu(2023)]%
        {lee2023chatgpt}
\bibfield{author}{\bibinfo{person}{Jonna Lee} {and} \bibinfo{person}{Meryem~Yilmaz Soylu}.} \bibinfo{year}{2023}\natexlab{}.
\newblock \showarticletitle{ChatGPT and Assessment in Higher Education}.
\newblock  (\bibinfo{year}{2023}).
\newblock


\bibitem[Liang et~al\mbox{.}(2021)]%
        {liang_towards_2021}
\bibfield{author}{\bibinfo{person}{Paul~Pu Liang}, \bibinfo{person}{Chiyu Wu}, \bibinfo{person}{Louis-Philippe Morency}, {and} \bibinfo{person}{Ruslan Salakhutdinov}.} \bibinfo{year}{2021}\natexlab{}.
\newblock \showarticletitle{Towards understanding and mitigating social biases in language models}. In \bibinfo{booktitle}{\emph{International {Conference} on {Machine} {Learning}}}. \bibinfo{publisher}{PMLR}, \bibinfo{pages}{6565--6576}.
\newblock
\urldef\tempurl%
\url{http://proceedings.mlr.press/v139/liang21a.html}
\showURL{%
\tempurl}


\bibitem[Liu et~al\mbox{.}(2023)]%
        {liu2023your}
\bibfield{author}{\bibinfo{person}{Jiawei Liu}, \bibinfo{person}{Chunqiu~Steven Xia}, \bibinfo{person}{Yuyao Wang}, {and} \bibinfo{person}{Lingming Zhang}.} \bibinfo{year}{2023}\natexlab{}.
\newblock \showarticletitle{Is your code generated by chatgpt really correct? rigorous evaluation of large language models for code generation}.
\newblock \bibinfo{journal}{\emph{arXiv preprint arXiv:2305.01210}} (\bibinfo{year}{2023}).
\newblock


\bibitem[Lo(2023)]%
        {lo2023clear}
\bibfield{author}{\bibinfo{person}{Leo~S Lo}.} \bibinfo{year}{2023}\natexlab{}.
\newblock \showarticletitle{The CLEAR path: A framework for enhancing information literacy through prompt engineering}.
\newblock \bibinfo{journal}{\emph{The Journal of Academic Librarianship}} \bibinfo{volume}{49}, \bibinfo{number}{4} (\bibinfo{year}{2023}), \bibinfo{pages}{102720}.
\newblock


\bibitem[Loubier(2023)]%
        {loubier2023chatgpt}
\bibfield{author}{\bibinfo{person}{Michael Loubier}.} \bibinfo{year}{2023}\natexlab{}.
\newblock \bibinfo{title}{ChatGPT: A Good Computer Engineering Student?: An Experiment on its Ability to Answer Programming Questions from Exams}.
\newblock
\newblock


\bibitem[Macfarlane et~al\mbox{.}(2014)]%
        {macfarlane2014academic}
\bibfield{author}{\bibinfo{person}{Bruce Macfarlane}, \bibinfo{person}{Jingjing Zhang}, {and} \bibinfo{person}{Annie Pun}.} \bibinfo{year}{2014}\natexlab{}.
\newblock \showarticletitle{Academic integrity: a review of the literature}.
\newblock \bibinfo{journal}{\emph{Studies in higher education}} \bibinfo{volume}{39}, \bibinfo{number}{2} (\bibinfo{year}{2014}), \bibinfo{pages}{339--358}.
\newblock


\bibitem[MacNeil et~al\mbox{.}(2022)]%
        {macneil_automatically_2022}
\bibfield{author}{\bibinfo{person}{Stephen MacNeil}, \bibinfo{person}{Andrew Tran}, \bibinfo{person}{Dan Mogil}, \bibinfo{person}{Seth Bernstein}, \bibinfo{person}{Erin Ross}, {and} \bibinfo{person}{Ziheng Huang}.} \bibinfo{year}{2022}\natexlab{}.
\newblock \showarticletitle{Generating Diverse Code Explanations using the GPT-3 Large Language Model}. In \bibinfo{booktitle}{\emph{Proceedings of the 2022 ACM Conference on International Computing Education Research - Volume 2}} \emph{(\bibinfo{series}{ICER ’22}, Vol.~\bibinfo{volume}{2})}. \bibinfo{publisher}{Association for Computing Machinery}, \bibinfo{address}{New York, NY, USA}, \bibinfo{pages}{37–39}.
\newblock
\showISBNx{978-1-4503-9195-5}
\urldef\tempurl%
\url{https://doi.org/10.1145/3501709.3544280}
\showDOI{\tempurl}


\bibitem[Malinka et~al\mbox{.}(2023)]%
        {malinka2023educational}
\bibfield{author}{\bibinfo{person}{Kamil Malinka}, \bibinfo{person}{Martin Peres{\'\i}ni}, \bibinfo{person}{Anton Firc}, \bibinfo{person}{Ondrej Hujn{\'a}k}, {and} \bibinfo{person}{Filip Janus}.} \bibinfo{year}{2023}\natexlab{}.
\newblock \showarticletitle{On the educational impact of ChatGPT: Is Artificial Intelligence ready to obtain a university degree?}. In \bibinfo{booktitle}{\emph{Proceedings of the 2023 Conference on Innovation and Technology in Computer Science Education V. 1}}. \bibinfo{pages}{47--53}.
\newblock


\bibitem[Mischak et~al\mbox{.}(2023)]%
        {mischak2023challenges}
\bibfield{author}{\bibinfo{person}{Robert Mischak}, \bibinfo{person}{Robert Pucher}, {et~al\mbox{.}}} \bibinfo{year}{2023}\natexlab{}.
\newblock \showarticletitle{Challenges for Computer Science Education Arising from new AI Systems like ChatGPT}. In \bibinfo{booktitle}{\emph{Conference Proceedings. The Future of Education 2023}}.
\newblock


\bibitem[Moore et~al\mbox{.}(2023)]%
        {moore2023empowering}
\bibfield{author}{\bibinfo{person}{Steven Moore}, \bibinfo{person}{Richard Tong}, \bibinfo{person}{Anjali Singh}, \bibinfo{person}{Zitao Liu}, \bibinfo{person}{Xiangen Hu}, \bibinfo{person}{Yu Lu}, \bibinfo{person}{Joleen Liang}, \bibinfo{person}{Chen Cao}, \bibinfo{person}{Hassan Khosravi}, \bibinfo{person}{Paul Denny}, {et~al\mbox{.}}} \bibinfo{year}{2023}\natexlab{}.
\newblock \showarticletitle{Empowering education with llms-the next-gen interface and content generation}. In \bibinfo{booktitle}{\emph{International Conference on Artificial Intelligence in Education}}. Springer, \bibinfo{pages}{32--37}.
\newblock


\bibitem[Mvondo et~al\mbox{.}(2023)]%
        {mvondo2023generative}
\bibfield{author}{\bibinfo{person}{Gustave Florentin~Nkoulou Mvondo}, \bibinfo{person}{Ben Niu}, {and} \bibinfo{person}{Salman Eivazinezhad}.} \bibinfo{year}{2023}\natexlab{}.
\newblock \showarticletitle{Generative Conversational AI And Academic Integrity: A Mixed Method Investigation To Understand The Ethical Use of LLM Chatbots In Higher Education}.
\newblock \bibinfo{journal}{\emph{Available at SSRN 4548263}} (\bibinfo{year}{2023}).
\newblock


\bibitem[Newton and Xiromeriti(2023)]%
        {newton2023chatgpt}
\bibfield{author}{\bibinfo{person}{Philip~Mark Newton} {and} \bibinfo{person}{Maira Xiromeriti}.} \bibinfo{year}{2023}\natexlab{}.
\newblock \showarticletitle{ChatGPT performance on MCQ exams in higher education. A pragmatic scoping review}.
\newblock  (\bibinfo{year}{2023}).
\newblock


\bibitem[Ngo(2023)]%
        {ngo2023perception}
\bibfield{author}{\bibinfo{person}{Thi Thuy~An Ngo}.} \bibinfo{year}{2023}\natexlab{}.
\newblock \showarticletitle{The Perception by University Students of the Use of ChatGPT in Education}.
\newblock \bibinfo{journal}{\emph{International Journal of Emerging Technologies in Learning (Online)}} \bibinfo{volume}{18}, \bibinfo{number}{17} (\bibinfo{year}{2023}), \bibinfo{pages}{4}.
\newblock


\bibitem[Onal and Kulavuz-Onal(2023)]%
        {onal2023cross}
\bibfield{author}{\bibinfo{person}{Sinan Onal} {and} \bibinfo{person}{Derya Kulavuz-Onal}.} \bibinfo{year}{2023}\natexlab{}.
\newblock \showarticletitle{A Cross-Disciplinary Examination of the Instructional Uses of ChatGPT in Higher Education}.
\newblock \bibinfo{journal}{\emph{Journal of Educational Technology Systems}} (\bibinfo{year}{2023}), \bibinfo{pages}{00472395231196532}.
\newblock


\bibitem[Orenstrakh et~al\mbox{.}(2023)]%
        {orenstrakh2023detecting}
\bibfield{author}{\bibinfo{person}{Michael~Sheinman Orenstrakh}, \bibinfo{person}{Oscar Karnalim}, \bibinfo{person}{Carlos~Anibal Suarez}, {and} \bibinfo{person}{Michael Liut}.} \bibinfo{year}{2023}\natexlab{}.
\newblock \showarticletitle{Detecting llm-generated text in computing education: A comparative study for chatgpt cases}.
\newblock \bibinfo{journal}{\emph{arXiv preprint arXiv:2307.07411}} (\bibinfo{year}{2023}).
\newblock


\bibitem[Perera and Lankathilaka(2023)]%
        {perera2023ai}
\bibfield{author}{\bibinfo{person}{Pethigmage Perera} {and} \bibinfo{person}{Madushan Lankathilaka}.} \bibinfo{year}{2023}\natexlab{}.
\newblock \showarticletitle{AI in higher education: A literature review of chatgpt and guidelines for responsible implementation}.
\newblock \bibinfo{journal}{\emph{International Journal of Research and Innovation in Social Science}} \bibinfo{volume}{7}, \bibinfo{number}{6} (\bibinfo{year}{2023}), \bibinfo{pages}{306--314}.
\newblock


\bibitem[Qadir(2022)]%
        {qadir_engineering_2022}
\bibfield{author}{\bibinfo{person}{Junaid Qadir}.} \bibinfo{year}{2022}\natexlab{}.
\newblock \showarticletitle{Engineering Education in the Era of ChatGPT: Promise and Pitfalls of Generative AI for Education}.
\newblock  (\bibinfo{date}{Dec.} \bibinfo{year}{2022}).
\newblock
\urldef\tempurl%
\url{https://doi.org/10.36227/techrxiv.21789434.v1}
\showDOI{\tempurl}


\bibitem[Qureshi(2023)]%
        {qureshi2023exploring}
\bibfield{author}{\bibinfo{person}{Basit Qureshi}.} \bibinfo{year}{2023}\natexlab{}.
\newblock \showarticletitle{Exploring the use of chatgpt as a tool for learning and assessment in undergraduate computer science curriculum: Opportunities and challenges}.
\newblock \bibinfo{journal}{\emph{arXiv preprint arXiv:2304.11214}} (\bibinfo{year}{2023}).
\newblock


\bibitem[Rajala et~al\mbox{.}(2023)]%
        {rajala2023call}
\bibfield{author}{\bibinfo{person}{Jaakko Rajala}, \bibinfo{person}{Jenni Hukkanen}, \bibinfo{person}{Maria Hartikainen}, {and} \bibinfo{person}{Pia Niemel{\"a}}.} \bibinfo{year}{2023}\natexlab{}.
\newblock \showarticletitle{$\backslash$" Call me Kiran$\backslash$"--ChatGPT as a Tutoring Chatbot in a Computer Science Course}. In \bibinfo{booktitle}{\emph{Proceedings of the 26th International Academic Mindtrek Conference}}. \bibinfo{pages}{83--94}.
\newblock


\bibitem[Reiche and Leidner(2023)]%
        {reiche2023bridging}
\bibfield{author}{\bibinfo{person}{Michael Reiche} {and} \bibinfo{person}{J Leidner}.} \bibinfo{year}{2023}\natexlab{}.
\newblock \showarticletitle{Bridging the Programming Skill Gap with ChatGPT: A Machine Learning Project with Business Students}.
\newblock \bibinfo{journal}{\emph{Nowacyk et al., S.(ed.) ECAI}} (\bibinfo{year}{2023}).
\newblock


\bibitem[Reynolds and McDonell(2021)]%
        {Reynolds_McDonell_2021}
\bibfield{author}{\bibinfo{person}{Laria Reynolds} {and} \bibinfo{person}{Kyle McDonell}.} \bibinfo{year}{2021}\natexlab{}.
\newblock \showarticletitle{Prompt Programming for Large Language Models: Beyond the Few-Shot Paradigm}.
\newblock  \bibinfo{number}{arXiv:2102.07350} (\bibinfo{date}{Feb.} \bibinfo{year}{2021}).
\newblock
\urldef\tempurl%
\url{http://arxiv.org/abs/2102.07350}
\showURL{%
\tempurl}
\newblock
\shownote{arXiv:2102.07350 [cs]}.


\bibitem[Rudolph et~al\mbox{.}(2023a)]%
        {rudolph2023chatgpt}
\bibfield{author}{\bibinfo{person}{J{\"u}rgen Rudolph}, \bibinfo{person}{Samson Tan}, {and} \bibinfo{person}{Shannon Tan}.} \bibinfo{year}{2023}\natexlab{a}.
\newblock \showarticletitle{ChatGPT: Bullshit spewer or the end of traditional assessments in higher education?}
\newblock \bibinfo{journal}{\emph{Journal of Applied Learning and Teaching}} \bibinfo{volume}{6}, \bibinfo{number}{1} (\bibinfo{year}{2023}).
\newblock


\bibitem[Rudolph et~al\mbox{.}(2023b)]%
        {rudolph2023war}
\bibfield{author}{\bibinfo{person}{J{\"u}rgen Rudolph}, \bibinfo{person}{Shannon Tan}, {and} \bibinfo{person}{Samson Tan}.} \bibinfo{year}{2023}\natexlab{b}.
\newblock \showarticletitle{War of the chatbots: Bard, Bing Chat, ChatGPT, Ernie and beyond. The new AI gold rush and its impact on higher education}.
\newblock \bibinfo{journal}{\emph{Journal of Applied Learning and Teaching}} \bibinfo{volume}{6}, \bibinfo{number}{1} (\bibinfo{year}{2023}).
\newblock


\bibitem[Sanfilippo et~al\mbox{.}(2023)]%
        {sanfilippo2023privacy}
\bibfield{author}{\bibinfo{person}{Madelyn~Rose Sanfilippo}, \bibinfo{person}{Noah Apthorpe}, \bibinfo{person}{Karoline Brehm}, {and} \bibinfo{person}{Yan Shvartzshnaider}.} \bibinfo{year}{2023}\natexlab{}.
\newblock \showarticletitle{Privacy governance not included: analysis of third parties in learning management systems}.
\newblock \bibinfo{journal}{\emph{Information and Learning Sciences}} \bibinfo{volume}{124}, \bibinfo{number}{9/10} (\bibinfo{year}{2023}), \bibinfo{pages}{326--348}.
\newblock


\bibitem[Shaengchart et~al\mbox{.}(2023)]%
        {shaengchart2023factors}
\bibfield{author}{\bibinfo{person}{Yarnaphat Shaengchart}, \bibinfo{person}{Nalinpat Bhumpenpein}, \bibinfo{person}{Kett Kongnakorn}, \bibinfo{person}{Phanuwit Khwannu}, \bibinfo{person}{Apisit Tiwtakul}, {and} \bibinfo{person}{Surachai Detmee}.} \bibinfo{year}{2023}\natexlab{}.
\newblock \showarticletitle{Factors influencing the acceptance of ChatGPT usage among higher education students in Bangkok, Thailand}.
\newblock \bibinfo{journal}{\emph{Advance Knowledge for Executives}} \bibinfo{volume}{2}, \bibinfo{number}{4} (\bibinfo{year}{2023}), \bibinfo{pages}{1--14}.
\newblock


\bibitem[Shardlow and Latham(2023)]%
        {shardlow2023chatgpt}
\bibfield{author}{\bibinfo{person}{Matthew Shardlow} {and} \bibinfo{person}{Annabel Latham}.} \bibinfo{year}{2023}\natexlab{}.
\newblock \showarticletitle{ChatGPT in computing education: a policy whitepaper}.
\newblock  (\bibinfo{year}{2023}).
\newblock


\bibitem[Singh et~al\mbox{.}(2023)]%
        {singh2023exploring}
\bibfield{author}{\bibinfo{person}{Harpreet Singh}, \bibinfo{person}{Mohammad-Hassan Tayarani-Najaran}, {and} \bibinfo{person}{Muhammad Yaqoob}.} \bibinfo{year}{2023}\natexlab{}.
\newblock \showarticletitle{Exploring Computer Science Students’ Perception of ChatGPT in Higher Education: A Descriptive and Correlation Study}.
\newblock \bibinfo{journal}{\emph{Education Sciences}} \bibinfo{volume}{13}, \bibinfo{number}{9} (\bibinfo{year}{2023}), \bibinfo{pages}{924}.
\newblock


\bibitem[Strzelecki(2023)]%
        {strzelecki2023use}
\bibfield{author}{\bibinfo{person}{Artur Strzelecki}.} \bibinfo{year}{2023}\natexlab{}.
\newblock \showarticletitle{To use or not to use ChatGPT in higher education? A study of students’ acceptance and use of technology}.
\newblock \bibinfo{journal}{\emph{Interactive Learning Environments}} (\bibinfo{year}{2023}), \bibinfo{pages}{1--14}.
\newblock


\bibitem[Sullivan et~al\mbox{.}(2023)]%
        {sullivan2023chatgpt}
\bibfield{author}{\bibinfo{person}{Miriam Sullivan}, \bibinfo{person}{Andrew Kelly}, {and} \bibinfo{person}{Paul McLaughlan}.} \bibinfo{year}{2023}\natexlab{}.
\newblock \showarticletitle{ChatGPT in higher education: Considerations for academic integrity and student learning}.
\newblock  (\bibinfo{year}{2023}).
\newblock


\bibitem[Tai and Chen(2023)]%
        {Tai_Chen_2023}
\bibfield{author}{\bibinfo{person}{Tzu-Yu Tai} {and} \bibinfo{person}{Howard Hao-Jan Chen}.} \bibinfo{year}{2023}\natexlab{}.
\newblock \showarticletitle{The impact of Google Assistant on adolescent EFL learners’ willingness to communicate}.
\newblock \bibinfo{journal}{\emph{Interactive Learning Environments}} \bibinfo{volume}{31}, \bibinfo{number}{3} (\bibinfo{date}{April} \bibinfo{year}{2023}), \bibinfo{pages}{1485–1502}.
\newblock
\showISSN{1049-4820}
\urldef\tempurl%
\url{https://doi.org/10.1080/10494820.2020.1841801}
\showDOI{\tempurl}


\bibitem[Tajik and Tajik(2023)]%
        {tajik2023comprehensive}
\bibfield{author}{\bibinfo{person}{Elham Tajik} {and} \bibinfo{person}{Fatemeh Tajik}.} \bibinfo{year}{2023}\natexlab{}.
\newblock \showarticletitle{A comprehensive Examination of the potential application of Chat GPT in Higher Education Institutions}.
\newblock \bibinfo{journal}{\emph{TechRxiv. Preprint}} (\bibinfo{year}{2023}), \bibinfo{pages}{1--10}.
\newblock


\bibitem[Vargas-Murillo et~al\mbox{.}(2023)]%
        {vargas2023challenges}
\bibfield{author}{\bibinfo{person}{Alfonso~Renato Vargas-Murillo}, \bibinfo{person}{Ilda Nadia~Monica de~la Asuncion}, \bibinfo{person}{Francisco de Jes{\'u}s Guevara-Soto}, {et~al\mbox{.}}} \bibinfo{year}{2023}\natexlab{}.
\newblock \showarticletitle{Challenges and Opportunities of AI-Assisted Learning: A Systematic Literature Review on the Impact of ChatGPT Usage in Higher Education}.
\newblock \bibinfo{journal}{\emph{International Journal of Learning, Teaching and Educational Research}} \bibinfo{volume}{22}, \bibinfo{number}{7} (\bibinfo{year}{2023}), \bibinfo{pages}{122--135}.
\newblock


\bibitem[Waits and Pomerantz(1997)]%
        {waits1997role}
\bibfield{author}{\bibinfo{person}{Bert Waits} {and} \bibinfo{person}{H Pomerantz}.} \bibinfo{year}{1997}\natexlab{}.
\newblock \showarticletitle{The Role of Calculators in Math Education}.
\newblock \bibinfo{journal}{\emph{USA: Department of Mathematics of The Ohio State University}} (\bibinfo{year}{1997}), \bibinfo{pages}{39--43}.
\newblock


\bibitem[Wang(2023)]%
        {wang2023navigating}
\bibfield{author}{\bibinfo{person}{Tianchong Wang}.} \bibinfo{year}{2023}\natexlab{}.
\newblock \showarticletitle{Navigating Generative AI (ChatGPT) in Higher Education: Opportunities and Challenges}. In \bibinfo{booktitle}{\emph{International Conference on Smart Learning Environments}}. Springer, \bibinfo{pages}{215--225}.
\newblock


\bibitem[Wang et~al\mbox{.}(2023)]%
        {wang2023exploring}
\bibfield{author}{\bibinfo{person}{Tianjia Wang}, \bibinfo{person}{Daniel~Vargas D{\'\i}az}, \bibinfo{person}{Chris Brown}, {and} \bibinfo{person}{Yan Chen}.} \bibinfo{year}{2023}\natexlab{}.
\newblock \showarticletitle{Exploring the Role of AI Assistants in Computer Science Education: Methods, Implications, and Instructor Perspectives}. In \bibinfo{booktitle}{\emph{2023 IEEE Symposium on Visual Languages and Human-Centric Computing (VL/HCC)}}. IEEE, \bibinfo{pages}{92--102}.
\newblock


\bibitem[Wen et~al\mbox{.}(2023)]%
        {wen2023empowering}
\bibfield{author}{\bibinfo{person}{Hao Wen}, \bibinfo{person}{Yuanchun Li}, \bibinfo{person}{Guohong Liu}, \bibinfo{person}{Shanhui Zhao}, \bibinfo{person}{Tao Yu}, \bibinfo{person}{Toby Jia-Jun Li}, \bibinfo{person}{Shiqi Jiang}, \bibinfo{person}{Yunhao Liu}, \bibinfo{person}{Yaqin Zhang}, {and} \bibinfo{person}{Yunxin Liu}.} \bibinfo{year}{2023}\natexlab{}.
\newblock \showarticletitle{Empowering llm to use smartphone for intelligent task automation}.
\newblock \bibinfo{journal}{\emph{arXiv preprint arXiv:2308.15272}} (\bibinfo{year}{2023}).
\newblock


\bibitem[Yilmaz and Yilmaz(2023)]%
        {yilmaz2023augmented}
\bibfield{author}{\bibinfo{person}{Ramazan Yilmaz} {and} \bibinfo{person}{Fatma Gizem~Karaoglan Yilmaz}.} \bibinfo{year}{2023}\natexlab{}.
\newblock \showarticletitle{Augmented intelligence in programming learning: Examining student views on the use of ChatGPT for programming learning}.
\newblock \bibinfo{journal}{\emph{Computers in Human Behavior: Artificial Humans}} \bibinfo{volume}{1}, \bibinfo{number}{2} (\bibinfo{year}{2023}), \bibinfo{pages}{100005}.
\newblock


\bibitem[Zhang et~al\mbox{.}(2023)]%
        {zhang2023s}
\bibfield{author}{\bibinfo{person}{Zhiping Zhang}, \bibinfo{person}{Michelle Jia}, \bibinfo{person}{Bingsheng Yao}, \bibinfo{person}{Sauvik Das}, \bibinfo{person}{Ada Lerner}, \bibinfo{person}{Dakuo Wang}, \bibinfo{person}{Tianshi Li}, {et~al\mbox{.}}} \bibinfo{year}{2023}\natexlab{}.
\newblock \showarticletitle{" It's a Fair Game'', or Is It? Examining How Users Navigate Disclosure Risks and Benefits When Using LLM-Based Conversational Agents}.
\newblock \bibinfo{journal}{\emph{arXiv preprint arXiv:2309.11653}} (\bibinfo{year}{2023}).
\newblock


\bibitem[Zhao et~al\mbox{.}(2023)]%
        {zhao2023survey}
\bibfield{author}{\bibinfo{person}{Wayne~Xin Zhao}, \bibinfo{person}{Kun Zhou}, \bibinfo{person}{Junyi Li}, \bibinfo{person}{Tianyi Tang}, \bibinfo{person}{Xiaolei Wang}, \bibinfo{person}{Yupeng Hou}, \bibinfo{person}{Yingqian Min}, \bibinfo{person}{Beichen Zhang}, \bibinfo{person}{Junjie Zhang}, \bibinfo{person}{Zican Dong}, {et~al\mbox{.}}} \bibinfo{year}{2023}\natexlab{}.
\newblock \showarticletitle{A survey of large language models}.
\newblock \bibinfo{journal}{\emph{arXiv preprint arXiv:2303.18223}} (\bibinfo{year}{2023}).
\newblock


\bibitem[Zhou et~al\mbox{.}(2023a)]%
        {zhou2023m}
\bibfield{author}{\bibinfo{person}{Kyrie~Zhixuan Zhou}, \bibinfo{person}{Jiaxun Cao}, \bibinfo{person}{Xiaowen Yuan}, \bibinfo{person}{Daniel~E Weissglass}, \bibinfo{person}{Zachary Kilhoffer}, \bibinfo{person}{Madelyn~Rose Sanfilippo}, {and} \bibinfo{person}{Xin Tong}.} \bibinfo{year}{2023}\natexlab{a}.
\newblock \showarticletitle{" I'm Not Confident in Debiasing AI Systems Since I Know Too Little": Teaching AI Creators About Gender Bias Through Hands-on Tutorials}.
\newblock \bibinfo{journal}{\emph{arXiv preprint arXiv:2309.08121}} (\bibinfo{year}{2023}).
\newblock


\bibitem[Zhou et~al\mbox{.}(2023b)]%
        {zhou2023toward}
\bibfield{author}{\bibinfo{person}{Kyrie~Zhixuan Zhou}, \bibinfo{person}{Madelyn Sanfilippo}, {and} \bibinfo{person}{Allison Sinnott}.} \bibinfo{year}{2023}\natexlab{b}.
\newblock \showarticletitle{Toward Ethical Use of Generative AI in AP Courses}.
\newblock \bibinfo{journal}{\emph{Information Matters}} \bibinfo{volume}{3}, \bibinfo{number}{11} (\bibinfo{year}{2023}).
\newblock


\bibitem[Zhou and Sanfilippo(2023)]%
        {zhou2023public}
\bibfield{author}{\bibinfo{person}{Kyrie~Zhixuan Zhou} {and} \bibinfo{person}{Madelyn~Rose Sanfilippo}.} \bibinfo{year}{2023}\natexlab{}.
\newblock \showarticletitle{Public perceptions of gender bias in large language models: Cases of chatgpt and ernie}.
\newblock \bibinfo{journal}{\emph{arXiv preprint arXiv:2309.09120}} (\bibinfo{year}{2023}).
\newblock


\bibitem[Zhou et~al\mbox{.}(2022)]%
        {zhou2022moral}
\bibfield{author}{\bibinfo{person}{Zhixuan Zhou}, \bibinfo{person}{Jiao Sun}, \bibinfo{person}{Jiaxin Pei}, \bibinfo{person}{Nanyun Peng}, {and} \bibinfo{person}{Jinjun Xiong}.} \bibinfo{year}{2022}\natexlab{}.
\newblock \showarticletitle{A Moral-and Event-Centric Inspection of Gender Bias in Fairy Tales at A Large Scale}.
\newblock \bibinfo{journal}{\emph{arXiv preprint arXiv:2211.14358}} (\bibinfo{year}{2022}).
\newblock


\end{thebibliography}
